\documentclass{aa}  
\usepackage{graphicx}
\usepackage[dvipsnames]{xcolor}
\usepackage{txfonts}
\usepackage{hyperref}
\usepackage{subcaption}
\usepackage{mathtools}

\definecolor{mygreen}{RGB}{0,153,76}

\usepackage[normalem]{ulem}
\usepackage{xcolor}
\definecolor{myteal}{HTML}{17A398}

\newcommand{\Htwo}[0]{\ensuremath{\mathrm{H}_2}}%

\begin{document} 

\title{Cosmic rays cannot explain the high ionisation rates in the Galactic centre}

\titlerunning{Cosmic rays cannot explain the high $\zeta$ in the GC}

\author{S. Ravikularaman \inst{1},
        S. Recchia \inst{2},
        V. H. M. Phan \inst{3}
        \and
        S. Gabici \inst{1}
    }
\authorrunning{S. Ravikularaman et al.}

   \institute{Université Paris Cité, CNRS, Astroparticule et Cosmologie, F-75013 Paris, France\\ \email{ravikularaman@apc.in2p3.fr}
   \and 
   INAF-Osservatorio Astronomico di Brera, Via Bianchi 46, I-23807 Merate, Italy
   \and
   Sorbonne Université, Observatoire de Paris, PSL Research University, LERMA, CNRS UMR 8112, 75005 Paris, France
   }
   
%\date{Received September 15, 1996; accepted March 16, 1997}

\abstract
{The $\Htwo$ ionisation rate in the central molecular zone, located in the Galactic centre, is estimated to be $\zeta\sim2\times10^{-14}~\mathrm{s}^{-1}$, based on observations of H$_3^+$ lines. This value is two to three orders of magnitude larger than that measured anywhere else in the Galaxy.}
{Due to the high density of the gas in the central molecular zone, UV and X-ray photons do not penetrate this region. Hence, cosmic rays are expected to be the exclusive agents of ionisation. A high cosmic-ray density has been invoked to explain the unusually high ionisation rate. 
However, this excess is not seen in the $\gamma$-ray emission from this region, which is produced by high-energy cosmic rays. 
Therefore, an excess is expected only in the low-energy cosmic-ray spectrum. 
Here, we derive constraints on this hypothetical low-energy component in the cosmic-ray spectra, and we question its plausibility.}
{To do so, we numerically solved the cosmic-ray transport equation in the central molecular zone, considering spatial diffusion, advection in the Galactic wind, re-acceleration in the ambient turbulence, and energy losses due to interactions with matter and radiation in the interstellar medium.
We derived stationary solutions under the assumption that cosmic rays are continuously injected by a source located in the Galactic centre.
The high-energy component in the cosmic-ray spectrum was then fitted to available $\gamma$-ray and radio data, and a steep low-energy component was added to the cosmic-ray spectrum to explain the large ionisation rates.}
{We find that injection spectra of $p^{-7}$ for protons below $p_\mathrm{enh,p}c\simeq780~\mathrm{MeV}$ and $p^{-5.2}$ for electrons below $p_\mathrm{enh,e}c=1.5~\mathrm{GeV}$ are needed to reach the observed ionisation rates. This corresponds to a cosmic-ray power of the order of $\sim10^{40-41}~\mathrm{erg}\,\mathrm{s}^{-1}$ injected at the Galactic centre. Not only is this unrealistic, but it is also impossible to reproduce a constant ionisation rate across the region, as observations suggest.}
{We conclude that cosmic rays alone cannot explain the high ionisation rates in the Galactic centre.}


\maketitle

%-------------------------------------------------------------------
\section{Introduction}
\label{sec:intro}

At the centre of the Milky Way, located $\sim8.5~\mathrm{kpc}$ from the Solar System, lies the supermassive black hole (SMBH) Sagittarius A$^*$ (Sgr A$^*$). It is surrounded by the Galaxy's densest and most massive molecular clouds (MCs), forming an asymmetric ring-like structure of molecular hydrogen gas \citep{fer07}. This region is named the central molecular zone (CMZ) and occupies a roughly cylindrical volume that has a radius of $\approx200~\mathrm{pc}$ and a height of $\approx 100$~pc. 

The star formation rate in the CMZ is $\approx0.07_{-0.02}^{+0.08}~\mathrm{M}_\odot\,\mathrm{yr}^{-1}$, a number significantly lower than expected given the amount of dense gas in this region \citep{hen23}. 
Hence, it is vital to understand the process and requirements for star formation in this region. 
One of the critical ingredients for star formation is the gas ionisation rate, which measures the number of ionisations an atom or a molecule of hydrogen undergoes per unit of time. 
A few different definitions exist for this quantity \citep{neufeld2017}. In this paper, we refer to the ionisation rate as the production rate of H$_2^+$ ions per hydrogen molecule.

The ionisation potential of $\Htwo$ is $I=15.426~\mathrm{eV}$. Both cosmic rays (CRs) and photons (UV, X-rays) can ionise $\Htwo$:
\begin{eqnarray}
\Htwo + {\rm CR} &\rightarrow& \Htwo^+ + e^- + {\rm CR}, \\
\Htwo + \gamma &\rightarrow& \Htwo^+ + e^- 
,\end{eqnarray} 
but it is believed that the first reaction largely dominates the second one in MCs, as ionising photons cannot penetrate large column densities of gas \citep{mck89}.
Among CRs, those with particle energy in the sub-giga-electronvolt domain are believed to be most effective in ionising interstellar matter \citep[see e.g.][and references therein]{gab22}.

The ionisation of molecular hydrogen is followed by the extremely rapid ion-neutral reaction \citep{oka06}:
\begin{equation}
    \Htwo^+ + \Htwo \rightarrow {\rm H}_3^+ + {\rm H}
,\end{equation} leading to the production of H$_3^+$ ions. 
Other ionisation processes, such as double or dissociative ionisation, are negligible compared to these processes \citep{pad09} and will be ignored in the following. 

${\rm H}_3^+$ is destroyed through dissociative recombination, according to one of the two following reactions \citep{oka06}:

\begin{eqnarray}
    {\rm H}_3^+ + e^- &\rightarrow& {\rm H} + {\rm H} + {\rm H}, \\
    {\rm H}_3^+ + e^- &\rightarrow& \Htwo + {\rm H}
.\end{eqnarray}
The rate of ${\rm H}_3^+$ destruction through recombination per unit volume is
\begin{equation}
    r_\mathrm{des}^{{\rm H}_3^+} = k_\mathrm{e}n({\rm H}_3^+)n(e^-)
,\end{equation}
where $k_\mathrm{e}$ is the Langevin rate constant for this reaction determined in laboratory experiments \citep{mcc03}. For typical cloud temperatures of a few tens of degrees, $k_\mathrm{e}\approx10^{-7}~\mathrm{cm}^3\,\mathrm{s}^{-1}$. By balancing the rates of formation and destruction of $\rm H_3^+$, we obtain an expression for the ionisation rate:
\begin{equation}
    \zeta_\mathrm{CR}^{\Htwo} = \frac{2k_\mathrm{e}x_\mathrm{e}}{f_{\Htwo}}\frac{N({\rm H}_3^+)}{L}
    \label{zL}
,\end{equation} where $x_\mathrm{e}$ is the electron fraction, $f_{\Htwo}=2n(\Htwo)/n_{\rm H}$ is the fraction of molecular hydrogen, and $N({\rm H}_3^+)$ is the H$_3^+$ column density along the line of sight of length $L$.
The latter quantity is the most uncertain and is usually estimated by assuming that clouds are roughly spherical. The other quantities on the right side of the equation can be constrained from observations \cite[see e.g.][and references therein for the case of a diffuse cloud towards $\zeta$~Persei]{mcc03}.

The ionisation rate in the local interstellar medium, measured by Voyager I, has been found to be in the range of $1.51-1.64\times10^{-17}~\mathrm{s}^{-1}$ \citep{cum16}.
Observations towards various lines of sight in the Galaxy have shown that the ionisation rate is of the order of $\approx10^{-16}~\mathrm{s}^{-1}$ in diffuse clouds and $\approx10^{-17}~\mathrm{s}^{-1}$ in dense clouds, with a quite large dispersion around these values \citep{cas98, ind12, pad09, pha18}. 
These values are much larger (by a factor of 10 to 100) than expected if one makes the assumption that the CR intensity changes only very mildly across the Galaxy \citep{pha18}.
A possible solution to this puzzle is to assume that, in fact, the intensity of low-energy (ionising) CRs varies spatially, and is larger close to CR accelerators \citep{pha21, pha23}.

The discrepancy is even more striking in the region very close to the Galactic centre (GC). 
By analysing the absorption spectra of various stars, \citet{oka05} measured the ${\rm H_3^+}$ column density and concluded that the ionisation rate in the CMZ is $(2-7)\times10^{-15}~\mathrm{s}^{-1}$. Based on new sight lines from \citet{geb11}, \citet{got11} concluded that the ionisation rates in the CMZ were $>10^{-15}~\mathrm{s}^{-1}$, which was later corroborated by other infrared observations \citep{got14}. \citet{lep16} question the validity of the linear relation between $N({\rm H}_3^+)$ and $\zeta$ and suggest that it only holds up to a certain value of the ionisation rate. 
Using the Meudon photon-dominated region (PDR) code \citep{lep06}, they found that the ionisation rate in the GC must be $1-11\times10^{-14}~\mathrm{s}^{-1}$.
This value is in accordance with \citet{oka19}'s most recent analysis of $~30$ stellar spectra, in which they raise their initial estimate by an order of magnitude to $\zeta_\mathrm{CMZ}=2\times10^{-14}~\mathrm{s}^{-1}$. 

\citet{yus07} used another method to constrain the ionisation rate. They analysed $\gamma$-rays, radio synchrotron emissions, and the Iron K$\alpha$ line and concluded that along certain lines of sight the ionisation rate could be as high as $5\times10^{-13}~\mathrm{s}^{-1}$. Later, they reduced this upper limit to $10^{-14}~\mathrm{s}^{-1}$ \citep{yus13}.

Molecules like ${\rm H_3O^+}$ are produced in MCs as the result of chemical reaction chains initiated by the ionisation of molecular hydrogen. These molecules have also been used to constrain the ionisation rate in the vicinity of the GC. In particular, \citet{van06} analysed ${\rm H_3O^+}$ maps of the Sgr B2 region and found that an ionisation rate of $4\times10^{-16}~\mathrm{s}^{-1}$ reproduced the ${\rm H_3O^+/H_2O}$ ratio in the cloud envelope. Although this value is lower than early estimates,  a more recent Herschel survey of ${\rm H_3O^+}$ suggests higher ionisation rates of $>10^{-15}~\mathrm{s}^{-1}$ \citep{ind15}.

Most recently, observations of other molecules have also hinted at a high ionisation rate. \citet{gin16} observed the $J=3-2$ transition of para-formaldehyde (p-${\rm H_2CO}$) to estimate the gas temperature of the CMZ and set the upper limit for the ionisation rate as $10^{-14}~\mathrm{s}^{-1}$. The measured abundances of ${\rm PO^+}$ \citep{riv22} and ${\rm HCOS^+}$ \citep{san24} towards the G+0.693$-$0.027 MC in the CMZ also imply an ionisation rate of $10^{-15}-10^{-14}~\mathrm{s}^{-1}$. 

The values of the ionisation rate measured in the CMZ employing various techniques range, then, from $\approx 10^{-15}$ to $\approx 10^{-13}$~s$^{-1}$, which are much larger (by up to few orders of magnitude) than the values measured from diffuse and dense clouds in the Galaxy.
Although an increase in the CR density, and therefore in the CR ionisation rate, might be expected towards the centre of the Galaxy, the increase is predicted to be at most a few factors \citep[see e.g.][]{wol03}.
It follows that, in order to explain the extremely large ionisation rates seen in the CMZ, one should invoke an equally large excess in sub-giga-electronvolt CRs there, and that this excess is due to a new component in the CR spectrum that is unrelated to the CR background that pervades the entire Galaxy.

This is even more puzzling, as tracers of the spatial distribution of higher-energy (giga-electronvolt) CRs in the Galaxy, such as $\gamma$-rays produced in CR interactions with interstellar matter, do not reveal the presence of a significant excess in the CMZ \cite[see e.g. Fig.~9 in][and references therein]{gab22}.
This implies that a large excess must only be present in low-energy CRs.

Remarkably, the H.E.S.S. Collaboration reported on the detection of the CMZ in tera-electronvolt $\gamma$-rays.
The spatial morphology of the emission indicates that a source of CRs has to be present in the vicinity of Sgr A$^*$ and that such a source is most probably a continuous injector of CRs \citep{hess16}.
The excess in multi-tera-electronvolt CRs derived from such measurements is several factors, and indicates that a new component in the spectrum of high-energy CRs is indeed present in the CMZ.

In this paper, we study whether it is possible for CRs to cause the very large ionisation rates observed in the CMZ. 
In particular, we assume that a CR accelerator is present in the vicinity of the GC and that such an accelerator injects particles continuously into the CMZ.
Our goal is to constrain the spectral energy distribution and the rate at which CRs have to be injected in order to explain the measured ionisation rates.
We conclude that the power requirement to fit data is exceedingly large, and therefore the cause of the large ionisation rate has to be searched for elsewhere.

In the next section, we describe the model for CR transport in the CMZ, including all the fiducial parameters. In Sec.\ref{sec:gamma} and \ref{sec:radio}, we constrain the high-energy CR proton and electron spectra from diffuse $\gamma$-ray and radio observations of the CMZ, respectively. We extrapolate these high-energy spectra to lower energies and compute the ionisation rates in Sec.\ref{sec:ion}. As these values are smaller by several orders of magnitude than the measured ionisation rates in the CMZ, we add a steep power-law component in the injection spectrum of CRs, and we try steeper and steeper slopes until the predicted ionisation rates reach the observed values. After studying the spatial variation in this value across the entire CMZ, we discuss the uncertainty introduced by the unknown depth, $L$, into the CMZ of the stars used for ${\rm H_3^+}$ absorption measurements. Most importantly, we compute the power $W_\mathrm{CR}$ of CRs and their energy density, $n_\mathrm{CR}$, needed to explain the measurements of the ionisation rate. We show in Sec.\ref{sec:add} that the CR spectra suitable to reproduce the large ionisation rates contradict other observations of the CMZ, and we eventually conclude that CRs cannot explain the large ionisation rates (Sec.\ref{sec:ccl}).

%--------------------------------------------------------------------
\section{Cosmic-ray transport in the central molecular zone}
\label{sec:transport}

\begin{figure}[htbp]
\centering
\includegraphics[width=\hsize]{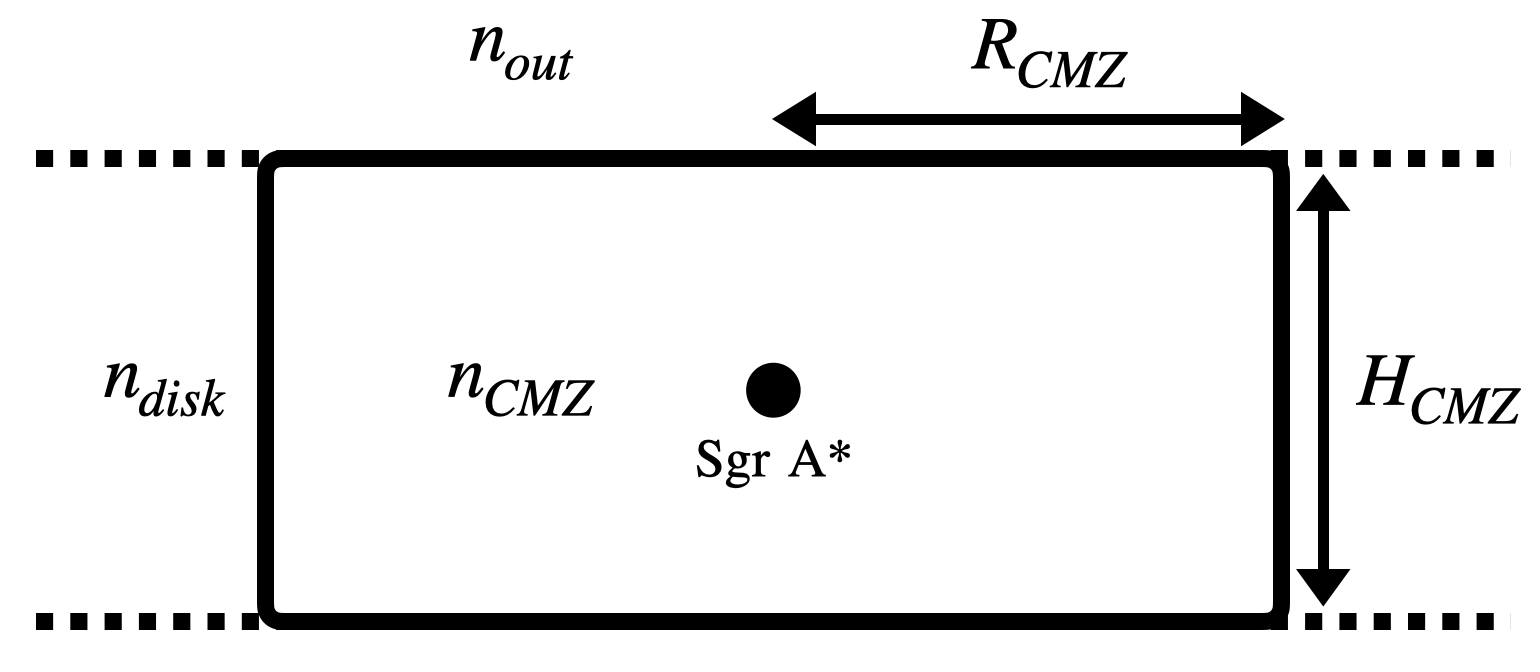}
\caption{Schematic edge-on view of the CMZ modelled as a cylinder of radius $R_\mathrm{CMZ}=200~\mathrm{pc}$ and height $H_\mathrm{CMZ}=100~\mathrm{pc}$, centred around Sgr A$^*$. The average density inside the CMZ is $n_\mathrm{CMZ}=1.5\times10^2~\mathrm{cm}^{-3}$. The densities in the Galactic disc and halo are taken as $n_\mathrm{disk}\sim1~\mathrm{cm}^{-3}$ and $n_\mathrm{out}\sim10^{-2}~\mathrm{cm}^{-3}$.}
\label{density}
\end{figure}

The CMZ was modelled as a cylinder of uniform gas density, $n_\mathrm{CMZ}$. 
The cylinder has a radius of $R_\mathrm{CMZ} \sim 200~\mathrm{pc}$ \citep{fer07} and height of $H_\mathrm{CMZ} \sim 100~\mathrm{pc}$, and is centred around Sgr~A$^*$ at $D_\mathrm{GC}\sim8.5~\mathrm{kpc}$ (see Fig. \ref{density}). The gas mass of the CMZ has been estimated to be $M_\mathrm{CMZ}\sim6\times10^{7}~\mathrm{M}_\odot$ \citep{dah98, tsu99}, which implies that the average gas density is $n_\mathrm{CMZ}=1.5\times10^2~\mathrm{cm}^{-3}$. 
The CMZ is embedded in the Galaxy's gaseous disc, which is assumed to have the same height as the CMZ, $H_\mathrm{CMZ}$, and characterised by a gas density of $n_\mathrm{disk}\sim1~\mathrm{cm}^{-3}$. 
%For $\vert{z}\vert > 50~\mathrm{pc}$, 
Outside of the Galactic disc, the gas density was set equal to  $n_\mathrm{out}\sim10^{-2}~\mathrm{cm}^{-3}$, which is appropriate to describe the Galactic halo.

The CMZ has been detected in very high-energy $\gamma$-rays \citep{hess06}.
The spatial correlation between the observed $\gamma$-rays and the gas distribution in the region suggests that such emission is due to the decay of neutral pions produced in inelastic interactions between CRs and ambient gas.
Moreover, a study of the morphology of the $\gamma$-ray emission suggests that the CRs responsible for such emission have been produced in the past by an accelerator located very close to the GC and that such an accelerator continuously injects relativistic particles into the surrounding medium \citep{hess16}.

After the escape from the accelerator, the transport of relativistic particles in and around the CMZ is dictated by spatial diffusion in the turbulent ambient magnetic field and advection in the Galactic wind.
At the same time, particles suffer energy losses due to interactions with ambient matter and radiation and are re-accelerated due to scattering off ambient magnetohydrodynamic waves. 

In order to describe the transport of CRs in the CMZ region, it is convenient to adopt a cylindrical set of co-ordinates centred on the GC, with the radial co-ordinate, $r$, directed along the Galactic plane and the height, $z$, orthogonal to it.
The equation describing the evolution in time, space, and momentum of the CR particle distribution function, $f(t,r,z,p)$, is \citep{ber90}:

\begin{equation}
\begin{split}
    \frac{\partial{f}}{\partial{t}} &= D_{rr}\frac{\partial^2{f}}{\partial{r^2}} + \left(\frac{D_{rr}}{r} + \frac{\partial{D_{rr}}}{\partial{r}}\right)\frac{\partial{f}}{\partial{r}} + D_{zz}\frac{\partial^2{f}}{\partial{z^2}} + \frac{\partial{D_{zz}}}{\partial{z}}\frac{\partial{f}}{\partial{z}} - v_\mathrm{w}\frac{\partial{f}}{\partial{z}}\\
    &+ \frac{1}{p^2}\frac{\partial}{\partial{p}}\left(p^2D_{pp}\frac{\partial{f}}{\partial{p}}\right) + \frac{p}{3}\frac{\partial{v_\mathrm{w}}}{\partial{z}}\frac{\partial{f}}{\partial{p}}  - \frac{1}{p^2}\frac{\partial}{\partial{p}}(\dot{p}p^2f) + Q
\end{split}
\label{eq:transport}
,\end{equation} 
where $D_{rr}$ and $D_{zz}$ are the CR spatial diffusion coefficients along $r$ and $z$, respectively, $v_\mathrm{w}$ is the Galactic wind velocity (assumed to be directed along the z axis), $D_{pp}$ the CR diffusion coefficient in momentum (particle re-acceleration), $\dot{p}$ the momentum loss rate, and $Q$ the particle injection rate.
In the following, we consider the case of an isotropic and spatially uniform diffusion ($D(p) = D_{rr} = D_{zz}$) and a wind speed that is independent of $z$.

\begin{table}
\caption{Physical parameters adopted to model the CMZ.}             
\label{tab:parameters}     
\centering                          
\begin{tabular}{c c c }        
\hline\hline      
 Parameter & Value \\
\hline
\hline 
   $D_\mathrm{GC}$ & $8.5~\mathrm{kpc}$ \\
   $R_\mathrm{CMZ}$ & $2.0\times10^{2}~\mathrm{pc}$ \\
   $H_\mathrm{CMZ}$ & $1.0\times10^{2}~\mathrm{pc}$ \\
   $V_\mathrm{CMZ}$ & $3.4\times10^{62}~\mathrm{cm}^{3}$ \\
   $M_\mathrm{CMZ}$ & $6\times10^{7}~\mathrm{M}_\odot$ \\
   $n_\mathrm{CMZ}$ & $1.5\times10^{2}~\mathrm{cm}^{-3}$ \\ 
   $n_\mathrm{disk}$ & $1.0~\mathrm{cm}^{-3}$ \\
   $n_\mathrm{out}$ & $1.0\times10^{-2}~\mathrm{cm}^{-3}$ \\
   $D(10~\mathrm{TeV})$ & $6\times10^{29}~\mathrm{cm}^2\,\mathrm{s}^{-1}$  \\  
   $\beta$ & $0.3$  \\
   $B_\mathrm{CMZ}$ & $1.5\times10^{2}~\mu{\rm G}$ \\ 
   $B_\mathrm{disk}$ & $3.0~\mu{\rm G}$ \\ 
   $B_\mathrm{out}$ & $3.0~\mu{\rm G}$ \\ 
   $v_\mathrm{A}$ & $1.0\times10^2~\mathrm{km}\,\mathrm{s}^{-1}$  \\
   $v_\mathrm{w}$ & $2.0\times10^2~\mathrm{km}\,\mathrm{s}^{-1}$  \\
   $T_\mathrm{NIR}$ ($\kappa_\mathrm{NIR}$) & $0.3~\mathrm{eV}$ ($1.3\times10^{-11}$) \\
   $T_\mathrm{FIR}$ ($\kappa_\mathrm{FIR}$) & $6\times10^{-3}~\mathrm{eV}$ ($9\times10^{-6}$) \\
   $T_\mathrm{CMB}$ ($\kappa_\mathrm{CMB}$) & $2.35\times10^{-4}~\mathrm{eV}$ ($0.99$) \\
\hline                          
\end{tabular}
\end{table}

The first four terms on the right-hand side of Eq.\ref{eq:transport} describe the spatial diffusion of particles, while the fifth term refers to advection in the Galactic wind. Diffusion in momentum space (re-acceleration) is modelled by the sixth term, while the seventh represents adiabatic losses. Finally, the eighth term accounts for momentum losses at a rate, $\dot{p}$, due to particle interaction with ambient matter and radiation.

\begin{figure*}[htbp]
\begin{subfigure}{.5\linewidth}
    \includegraphics[width=\linewidth]{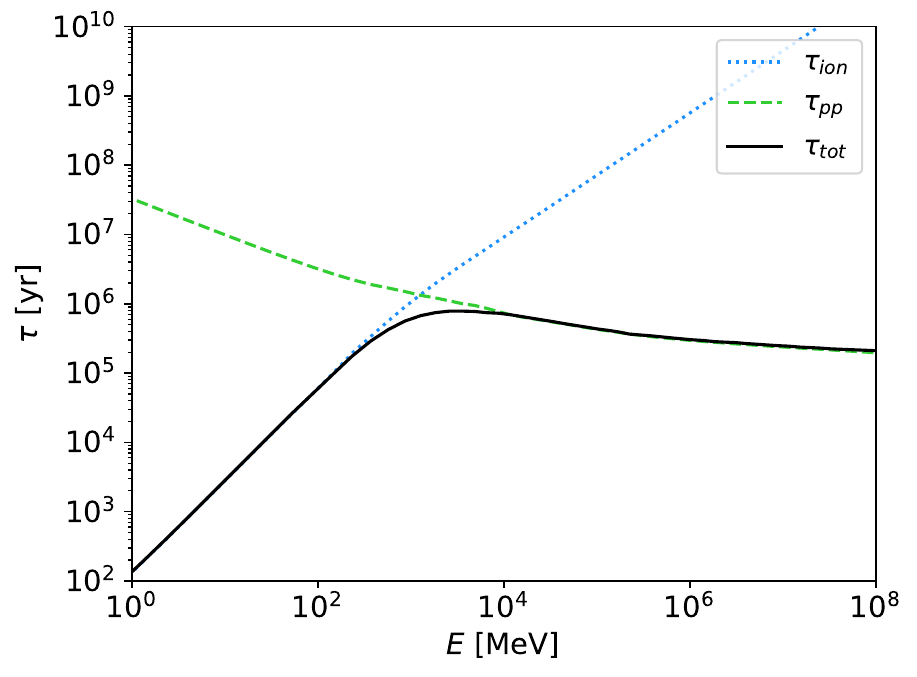}
\end{subfigure}
\hfill
\begin{subfigure}{.5\linewidth}
    \includegraphics[width=\linewidth]{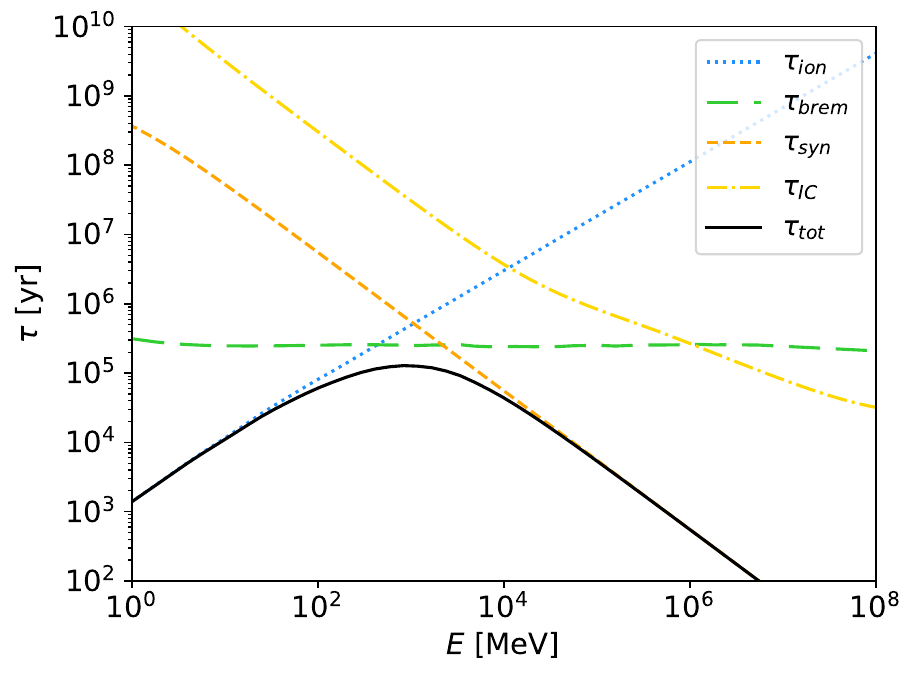}
\end{subfigure}
\caption{Characteristic timescales for CR energy losses in the CMZ (see Tab.~\ref{tab:parameters} to see what parameters were adopted). Proton loss times for ionisation and p-p interactions (left panel) and electron loss times for ionisation, synchrotron, bremsstrahlung, and inverse Compton scattering (right panel).}
\label{fig:losstimes}
\end{figure*}

The CR spatial diffusion coefficient is difficult to constrain from observations.
However, in order to have diffusive transport of CRs in the CMZ, the mean free path for spatial diffusion has to be much smaller than the size of the region, $R_\mathrm{CMZ}$.
This converts into an upper limit for the CR diffusion coefficient as $D \ll cR/3 \approx 6 \times10^{30}~\mathrm{cm}^2\,\mathrm{s}^{-1}$. Following \cite{hess16}, we assume the diffusion to be isotropic, and the diffusion coefficient is
\begin{equation}
    D_{rr} = D_{zz} = D(p) = 6\times10^{29}\left(\frac{pc}{10~\mathrm{TeV}}\right)^{\beta}~\mathrm{cm}^2\,\mathrm{s}^{-1}
,\end{equation} 
with $\beta=0.3$. 
Although theoretical considerations suggest that a break might appear in $D(p)$ in the low-energy (transrelativistic) regime \citep[see e.g.][and references therein]{pha21}, we adopt here a pure power law scaling down to arbitrarily low particle energies, as the presence of a break will not affect any of our results.
Our choice of $D$ makes the transport of CRs diffusive up to particle energies in the multi-peta-electronvolt domain.

The advection of CRs is caused by a Galactic wind with a discontinuous velocity at $z=0$ of the form $v(z) = \mathrm{sgn}(z)v_\mathrm{w}$. The value of the wind speed, $v_\mathrm{w}$, in the CMZ region is uncertain, with estimates ranging from $200~\mathrm{km}\,\mathrm{s}^{-1}$ to $1200~\mathrm{km}\,\mathrm{s}^{-1}$ \citep[see e.g.][]{cro11}.
To maximise the impact of CRs in the region, we adopt here the value corresponding to the lower bound in that interval:  $v_\mathrm{w}\sim200~\mathrm{km}\,\mathrm{s}^{-1}$.
Larger values of $v_\mathrm{w}$ would increase advection efficiency, resulting in lower densities of CRs in the CMZ region.

The CR momentum diffusion coefficient, $D_{pp}$, is expected to be inversely proportional to the spatial diffusion coefficient, as many repeated scatterings onto (moving) magnetic irregularities would enhance diffusion in momentum space while suppressing spatial diffusion.
If magnetic perturbances are Alfv\'en waves, the following heuristic expression can be adopted \citep{tho14}:
\begin{equation}
    D_{pp}(p) = \frac{p^2v^2_\mathrm{A}}{9D(p)}
,\end{equation} where $v_\mathrm{A}$ is the Alfv\'en velocity. 

The momentum loss rate, $\dot{p}$, accounts for energy loss processes due to interactions of CRs with ambient matter and radiation \citep[see][and references therein]{pad18,gab22}.
It describes ionisation losses at low particle energies, while at larger energies, bremsstrahlung, synchrotron, and inverse Compton losses dominate for CR electrons and proton-proton inelastic interactions for CR protons.

Ionisation, pion production, and bremsstrahlung losses depend only on the ambient gas density and are treated in the following as in \citet{pad18}. 
To describe inverse Compton losses, we followed \citet{kha14} and considered a target photon field with three thermal (grey body) components: near-infrared (NIR), far-infrared (FIR), and the cosmic microwave background (CMB), equal to the ones adopted by \citet{hin07} to model the radiation density in the CMZ region.
For synchrotron losses, the value of the ambient magnetic field, $B$, was kept as a free parameter, but its value was constrained to be larger than 50~$\mu$G, a lower limit derived observationally for the CMZ by \cite{cro10}. The magnetic field strength outside the CMZ was taken to be 5~$\mu$G. Although higher values have been suggested by \citet{orl13, planck16, orl19}, the choice of this value does not affect our results. The energy loss time for each mechanism, $l$, was then defined as
\begin{equation}
    \tau_{l}=E/\dot{E}
\label{eq:losstime}
,\end{equation} 
where $E$ is the particle kinetic energy and $\dot{E} = -{\rm d}E / {\rm d} t$ the energy loss rate.
The loss times are shown in Fig. \ref{fig:losstimes} for CR protons (left) and electrons (right).

Finally, the injection of CR particles was considered to be point-like in space and continuous in time.
Particle spectra were assumed to be power laws (or broken power laws, as in Sec.~\ref{sec:break}) in momentum, giving:
\begin{equation}
\label{eq:injection1}
    Q_\mathrm{i}(r, z, p) = Q_\mathrm{p,i}(p)\frac{\delta(r)}{2\pi{r}}\delta(z)
,\end{equation} where 
\begin{equation}
\label{eq:injection2}
    Q_\mathrm{p,i}(p) = Q_\mathrm{0,i}\left(\frac{pc}{10~\mathrm{TeV}}\right)^{-\delta_\mathrm{i}}
,\end{equation} where the subscript $i$ can refer to either protons ($i = p$) or electrons ($i = e$).

\begin{figure}[htbp]
\centering
\includegraphics[width=\hsize]{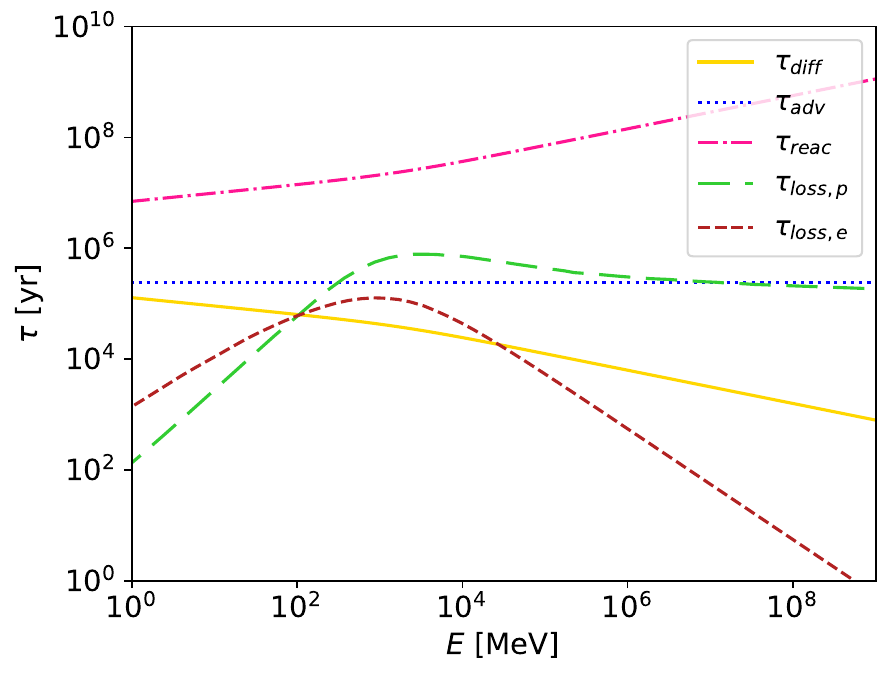}
\caption{Characteristic timescales for spatial diffusion, advection, re-acceleration, and losses for protons and electrons. The fiducial parameters are given in Table. \ref{tab:parameters}.}
\label{times}
\end{figure}

For each process affecting the propagation of CRs, it is possible to define a characteristic timescale based on the geometry of the CMZ and the typical physical parameters that characterise that region (see Tab.~\ref{tab:parameters}). This gives us an idea of the relative importance of a mechanism at any given energy (see Fig.~\ref{times}, where all relevant timescales are shown). 

The timescale for spatial diffusion was defined as the time needed for a CR of momentum $p$ to diffuse across a distance of $R_\mathrm{CMZ}$:
\begin{equation}
    \tau_\mathrm{diff}(p) = \frac{R_\mathrm{CMZ}^2}{6D(p)}  .
\end{equation} 
The advection timescale was defined as the time needed for a particle to be advected over a vertical distance of $H_\mathrm{CMZ}/2$:
\begin{equation}
    \tau_\mathrm{adv}(p) = \frac{H_\mathrm{CMZ}}{2v_\mathrm{w}} ~ .
\end{equation} 
The timescale for re-acceleration was defined as the time required for a particle to reach a momentum, $p$:
\begin{equation}
    \tau_\mathrm{reac}(p) = \frac{p^2}{6D_{pp}(p)} ~ . 
\end{equation} 
The ratio between the re-acceleration and diffusion timescale is
\begin{equation}
\frac{\tau_\mathrm{reac}}{\tau_\mathrm{diff}} = \frac{9 ~ D^2}{v_\mathrm{A}^2 R_\mathrm{CMZ}^2 } \sim 1.6 \times 10^3 \left( \frac{D}{10^{28}~{\rm cm^2/s}} \right)^2 \left( \frac{v_\mathrm{A}}{10~\rm km/s} \right)^{-2}
,\end{equation}
which is large for any plausible value of the Alfv\'en speed and of the diffusion coefficient. Indeed, re-acceleration is not a relevant process at any energy scale of our interest (see Fig.~\ref{times}).
Therefore, in the following, we have fixed $v_\mathrm{A} = 100~\mathrm{km}\,\mathrm{s}^{-1}$ both inside and outside the CMZ as this parameter does not significantly affect our results. It has been verified that adopting $v_\mathrm{A}=0~\mathrm{km}\,\mathrm{s}^{-1}$ does not give any observable differences in the CR spectra.
Finally, the loss timescale was calculated separately for protons and electrons using the loss times (Eq. \ref{eq:losstime}) for all the mechanisms considered.

The steady-state solution for a given injection spectrum was obtained by solving the transport equation numerically with a Crank-Nicholson scheme \citep{evo17, kis14, nr92} in which the delta-function in space describing particle injection (Eq.~\ref{eq:injection1}) was modelled as a narrow Gaussian. The physical parameters adopted to compute all the characteristic times are given in Table. \ref{tab:parameters}.

\section{$\gamma$-ray observations of the central molecular zone}
\label{sec:gamma}

The $\gamma$-ray emission from the CMZ has been observed by the High Energy Stereoscopic System (H.E.S.S.) \citep{hess06, hess16, hess18}, the Fermi Gamma-ray Space Telescope \citep{gag17}, the Very Energetic Radiation Imaging Telescope Array System (VERITAS) \citep{veritas21}, and the Major Atmospheric Gamma Imaging Cherenkov (MAGIC) Telescopes \citep{magic20}. 
In the following, we compare our predictions with the observations performed by Fermi in the region of Galactic co-ordinates $\vert{l}\vert<0.8^\circ$ and $\vert{b}\vert<0.3^\circ$ \citep{gag17} and by H.E.S.S. in the regions defined by $\vert{l}\vert<0.8^\circ$ and $\vert{b}\vert<0.3^\circ$ \citep{hess06} and $\vert{l}\vert<1^\circ$ and $\vert{b}\vert<0.3^\circ$ \citep{hess18}.

The $\gamma$-rays observed from the Galactic disc result from the interactions of CRs with matter and radiation in the interstellar medium. 
The dominant process of $\gamma$-ray production is the interaction of CR protons with nuclei in the ISM. 
Leptonic processes such as inverse Compton scattering and non-thermal bremsstrahlung might also contribute to the $\gamma$-ray emission (though they are expected to be subdominant in the CMZ).

The CMZ region is characterised by a very large gas density, and its $\gamma$-ray emission correlates quite well with the spatial distribution of interstellar matter, pointing quite unambiguously to a hadronic origin of the observed $\gamma$-rays \citep{hess06}.
A more detailed analysis of the H.E.S.S. data suggests that the $\gamma$-ray emission is produced by CR nuclei that escaped from an accelerator located very close to the GC, which has been injecting particles in the surrounding medium for an extended period of time.

In the remainder of this section, we make use of the CR transport code described in Sec.~\ref{sec:transport} and we estimate the $\gamma$-ray emission produced by CR nuclei and electrons that have been produced by such a source and that are now filling the CMZ.

In order to estimate the spectrum of $\gamma$-rays produced in hadronic proton-proton interactions, we make use of the parameterisations of cross-sections provided by \citet{kaf14}, including the nuclear enhancement factor that accounts for the presence of nuclei heavier than protons in both CRs and interstellar matter.

To model inverse Compton scattering, we considered interactions between relativistic electrons and soft photons present in the CMZ, following the approach presented in \citet{kha14}.
The thermal components of the radiation field in the CMZ are NIR, peaking at a photon energy of 0.3 eV, FIR at $6 \times 10^{-3}$ eV, and the CMB at $2.35 \times 10^{-4}$ eV \citep{hin07}. The respective energy densities are 9, 1, and 0.26 $\mathrm{eV}\,\mathrm{cm}^{-3}$. Each component was assumed to be characterised by a diluted black-body spectrum, with the dilution factor, $\kappa$, computed using the Stefan-Boltzmann law:

\begin{equation}
    \kappa = \frac{wc}{4\sigma_\mathrm{sb}T^4}
,\end{equation} where $w$ is the photon energy density, $c$ the speed of light, $\sigma_\mathrm{sb}$ the Stefan-Boltzmann constant, and $T$ the photon temperature.

Finally, bremsstrahlung radiation is produced when electrons decelerate in the Coulomb field of an ion. 
The emitted $\gamma$-ray radiation was modelled using the cross-section appropriate for interactions of energetic electrons with neutral molecular hydrogen given in \citet{sch02}. 

\begin{figure*}[htbp]
\begin{subfigure}{.5\linewidth}
    \includegraphics[width=\linewidth]{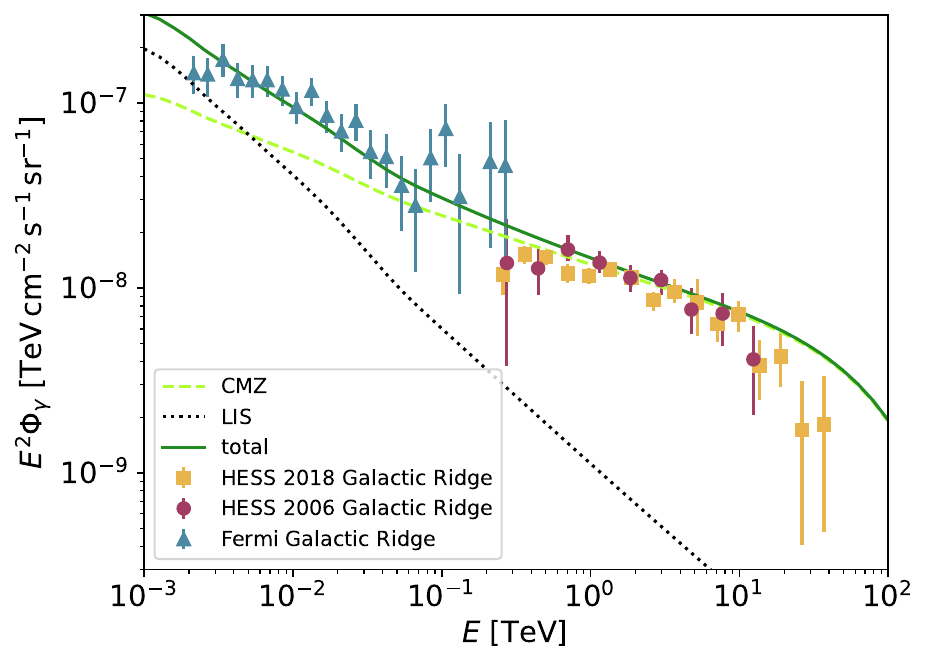}
\end{subfigure}
\hfill
\begin{subfigure}{.5\linewidth}
    \includegraphics[width=\linewidth]{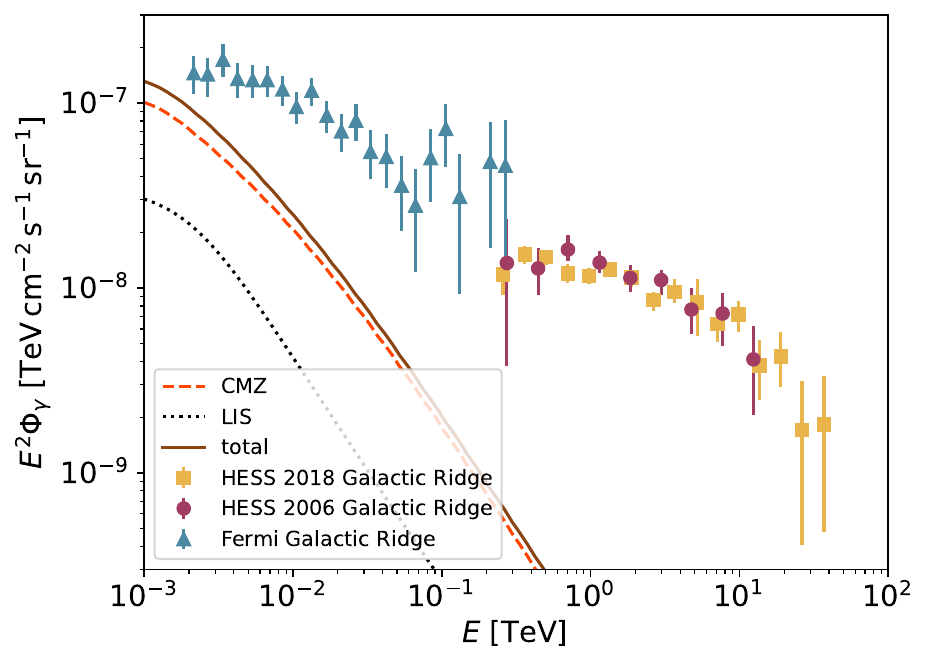}
\end{subfigure}
  \caption{$\gamma$-ray data for the CMZ region. Curves show theoretical results for hadronic (left panel) and leptonic (right panel) scenarios. Dashed lines show the emission from CRs accelerated within the CMZ (in a source located at the GC), while dotted lines show the contribution from ambient CRs, assumed to have a spectrum equal to the local interstellar one (LIS).}
\label{fig:gamma}
\end{figure*}

\subsection*{Derivation of cosmic-ray proton injection spectrum}

In this section, we make use of the CR transport code described above to model the propagation of CRs that escaped the accelerator located in the GC and now fill the CMZ region.
The time-dependent Eq.~\ref{eq:transport} was evolved in time until stationarity was reached. 
Then, the $\gamma$-ray emission from interactions of both CR nuclei and electrons in the ISM was computed. 
The spatial correlation between the $\gamma$-ray emission and the distribution of dense gas in the CMZ suggests a hadronic origin of the $\gamma$-ray photons \citep{hess16}. So, the fit to available data allows us to derive the injection spectrum for CR protons. 
We also consider and critically discuss leptonic scenarios in which the $\gamma$-ray data is used as an upper limit for the contribution from inverse Compton scattering and relativistic bremsstrahlung.

\begin{figure}[htbp]
\centering
\includegraphics[width=\hsize]{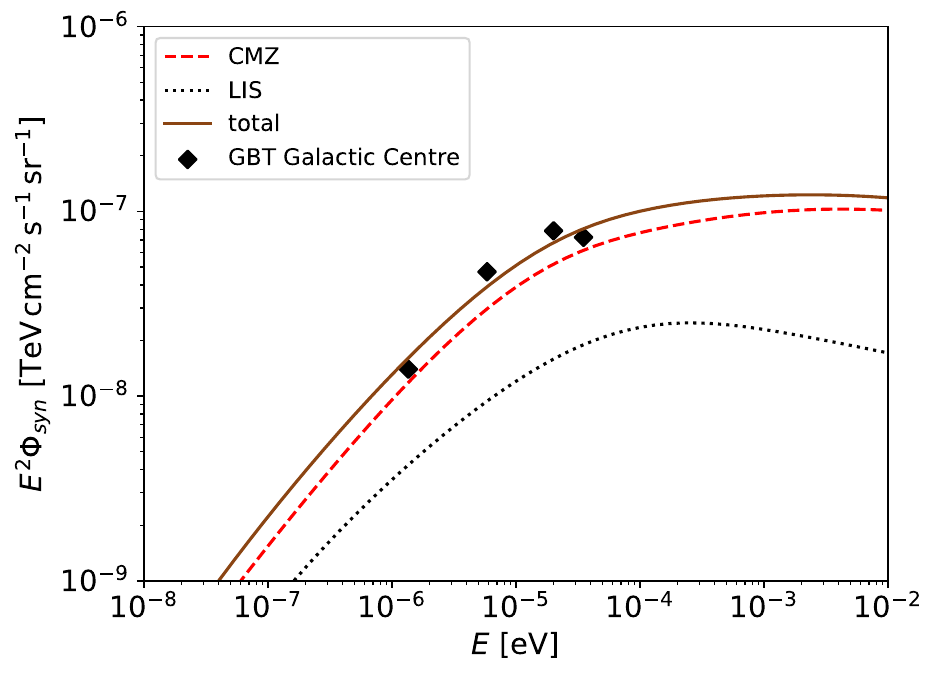}
\caption{Fit to radio data from the CMZ region. The solid line refers to the total predicted synchrotron emission from CR electrons. The dashed line shows the contribution from electrons accelerated inside the CMZ. The dotted line shows the contribution from the background sea of CR electrons (LIS) with a low-energy exponential cut-off to take into account the suppressed penetration of background CR electrons into the CMZ. We consider an average magnetic field strength of $B_\mathrm{CMZ}\sim150~\mu\mathrm{G}$.}
\label{fig:radio}
\end{figure}

The left and right panels of Fig.~\ref{fig:gamma} show the observed $\gamma$-ray spectrum of the CMZ region, together with curves showing the results of our hadronic and leptonic models obtained by solving Eq.~\ref{eq:transport}. Data points refer to Fermi and H.E.S.S. observations, as is indicated in the figure inset.
As we show in the same plot $\gamma$-ray brightnesses extracted from slightly different regions, we decided to compare them with a typical predicted brightness computed for the entire CMZ (model curves in the figure).
Such an approach, admittedly approximate, introduces uncertainty in our derived CR spectrum normalisation of the order of a few tens of percent.
This is not a concern, given that we are seeking to explain an excess of several orders of magnitude in the CR ionisation rate. We note that the same caveat applies to the fit to radio data presented in Fig.~\ref{fig:radio}.

The model curves shown in the figure refer to the emission from CR nuclei (left panel) and CR electrons (right panel).
The long dashed lines show our predictions for the $\gamma$-ray emission generated by CRs accelerated by the source located in the GC.
Dotted lines show the emission coming from the interactions of background Galactic CRs that are known to fill the entire Galactic disc.
As the spatial gradient of the latter component is believed to be very small, we assume that background CRs have the same intensity as the ones measured in the local interstellar medium (the label LIS stands for local interstellar spectrum) by Voyager and AMS-02 \citep[see e.g.][and references therein]{gab22}:

\begin{equation}
    J_{\rm LIS, p}(E_{\rm p}) = \frac{J_{\rm 0,p}\left(\frac{E_{\rm p}}{\mathrm{MeV}}\right)^{0.35}}{\left[1+\left(\frac{E}{80~\mathrm{MeV}}\right)^{1.3}\right]\left[1+\left(\frac{E_{\rm p}}{2.2~\mathrm{GeV}}\right)^{1.9/s}\right]^s}
,\end{equation} where $J_{\rm 0,p}=12.5~\mathrm{MeV}^{-1}\,\mathrm{m}^{-2}\,\mathrm{s}^{-1}\,\mathrm{sr}^{-1}$ and $s=2.1$.

The $\gamma$-ray emission from background CR nuclei dominates the flux in the giga-electronvolt energy domain, but it becomes subdominant in the tera-electronvolt one.
On the other hand, the emission from background CR electrons is always largely subdominant.

From Fig.~\ref{fig:gamma} (left panel), it can be seen that the hadronic scenario provides a good fit to $\gamma$-ray data if the injection spectrum of CR protons is characterised by:
\begin{eqnarray}
Q_\mathrm{p,p}(10~\mathrm{TeV}) &=& 1.1\times10^{29}~\mathrm{MeV}^{-1}~\mathrm{s}^{-1} \\
\delta_\mathrm{p} &=& 4.1
.\end{eqnarray}
This corresponds to a CR proton luminosity above particle energy of 10 TeV equal to $W_\mathrm{CR}(E_\mathrm{p}\geq 10~\mathrm{TeV}) \approx 7\times10^{37}~\mathrm{erg}\,\mathrm{s}^{-1}$, which is comparable with the early estimation from the H.E.S.S. Collaboration $\approx4\times10^{37}(D(10~{\rm TeV})/10^{30}~\mathrm{cm}^2\,\mathrm{s}^{-1})~\mathrm{erg}\,\mathrm{s}^{-1}$ \citep{hess16}. The leptonic scenario (right panel of Fig.~\ref{fig:gamma}) will be discussed in the next section.

\section{Radio observations of the central molecular zone}
\label{sec:radio}

The radio emission from the CMZ region has been mapped by the Very Large Array (VLA) \citep{lar05}, the Green Bank Telescope (GBT) \citep{law08}, and MeerKAT \citep{hey22}. 
We consider here the total continuum flux at 0.325, 1.40, 8.5, and 5$~\mathrm{GHz}$ obtained by \citet{law08}, \citet{yus13} from the inner $2^\circ\times 0.85^\circ$ of the GC region.
According to \citet{law08} and \citet{yus13}, the contribution from non-thermal emissions at these frequencies -- that is, synchrotron radiation from CR electrons -- is largely dominant.
Hence, the radio continuum emission can be used to probe the intensity of CR electrons close to the GC. 

\subsection*{Derivation of the cosmic-ray electron spectrum}
\label{sec:break}

Here, we constrain the spectrum of CR electrons in the CMZ from the radio observations of the inner from the GBT \citep{law08, yus13}. To fit this data, we assumed that CR electrons are injected at the central source with a spectrum that is a broken power law in momentum:
\begin{equation}
        Q_\mathrm{p,e}(p)= 
\begin{dcases}
    Q_*\left(\frac{p}{p_*}\right)^{-\delta_\mathrm{e,1}} ,& \text{if } p\leq p_*\\
    Q_{*}\left(\frac{p}{p_*}\right)^{-\delta_\mathrm{e,2}} ,& \text{otherwise}
\end{dcases}
\label{eq:bknpl}
,\end{equation} 
where $p_*c=1.5~\mathrm{GeV}$. 
Similarly to the case of $\gamma$-ray emissions, we propagated this injection spectrum using Eq.(\ref{eq:transport}) and the steady-state solution was used to compute synchrotron emissions.

The radio data can be fitted using a broken power law, as is defined in Eq.(\ref{eq:bknpl}) using the following normalisation:
\begin{equation}
    Q_\mathrm{p,e}(10~{\rm TeV}) = 1.5\times10^{27}~\mathrm{MeV}^{-1}~\mathrm{s}^{-1}
,\end{equation}
and spectral indices
\begin{eqnarray}
    \delta_\mathrm{e,1} &= 3.25\\
    \delta_\mathrm{e,2} &= 4.4 
,\end{eqnarray}
where $\delta_\mathrm{e,1}$ is mostly constrained by radio data, while $\delta_\mathrm{e,2}$ has been chosen to not overshoot the $\gamma$-rays. In this respect, the right panel of Fig.(\ref{fig:gamma}) shows the observed $\gamma$-ray spectrum of the CMZ plotted with the contributions expected from inverse Compton and relativistic bremsstrahlung produced by the steady-state spectrum obtained from Eq.(\ref{eq:bknpl}).

To the contribution from the CRs accelerated in the GC, we added the radio emission from the background Galactic CR electrons. The intensity of the background CR electrons, obtained from Voyager and AMS-02 data \citep[see e.g.][and references therein]{gab22}, is:

\begin{equation}
    J_{\rm LIS, e}(E_{\rm e}) = \frac{J_{\rm 0,e}\left(\frac{E}{\mathrm{MeV}}\right)^{-1.3}\exp{\left(-\frac{100~\mathrm{MeV}}{E_{\rm e}}\right)}}{\left[1+\left(\frac{E}{65~\mathrm{MeV}}\right)^{0.6/s_1}\right]^{s_1}\left[1+\left(\frac{E}{2.8~\mathrm{GeV}}\right)^{1.38/s_2}\right]^{s_2}}
,\end{equation} where $J_{\rm 0,e}=5\times10^{3}~\mathrm{MeV}^{-1}\,\mathrm{m}^{-2}\,\mathrm{s}^{-1}\,\mathrm{sr}^{-1}$, $s_1=0.2$ and $s_2=0.5$.

Before computing the synchrotron emission from the ambient (LIS) CRs, we added a low-energy exponential cut-off at an energy of 100 MeV.
This is the energy for which the CR electron loss time equals the diffusion time across the CMZ (see Fig.~\ref{times}), and the cut-off is intended to roughly mimic the effect of the penetration of Galactic CRs into the CMZ (see e.g. \cite{dog21}).

\section{Cosmic-ray ionisation rate}
\label{sec:ion}

Once we had imposed constraints on the intensity and spectral energy distribution of CR nuclei and electrons in the CMZ, we investigated the impact that these particles have on the gas in the CMZ. 
In particular, we computed the CR ionisation rate of ambient gas, and we compared that with available measurements of this parameter.
To do so, it was necessary to extrapolate to low energies the CR spectra derived above, as low-energy particles (roughly in the sub-giga-electronvolt domain) are believed to be the most effective ionisation agents in MCs.

As is described in Sec.~\ref{sec:intro}, the ionisation rate of H$_2$ molecules in dense MCs can be measured by means of observation of molecular lines such as, most notably, those of H$_3^+$ falling in the IR band \citep{oka06, mil20}.
This is because the production of H$_3^+$ in MCs follows in a straightforward way from the ionisation of molecular hydrogen.

\begin{figure}[htbp]
\centering
   \includegraphics[width=\hsize]{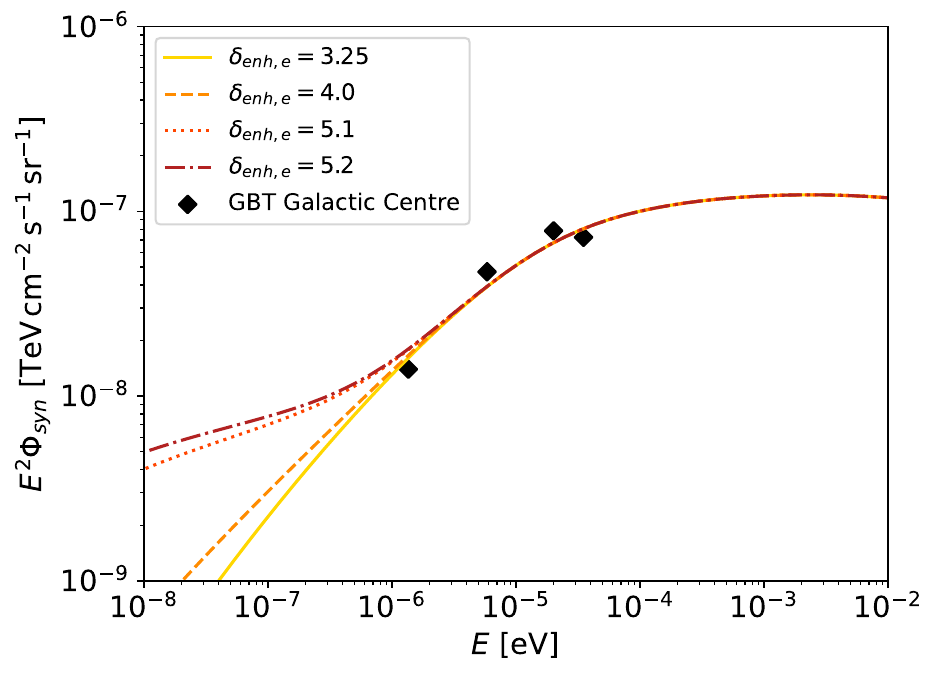}
     \caption{Expected synchrotron radio emission from the enhanced CR electron injection spectra considered in this work (see main text for details), compared to radio observations of the CMZ. }
     \label{fig:radiofs}
\end{figure}

If $f_\mathrm{p}$ and $f_\mathrm{e}$ are the CR particle distribution functions for protons and electrons, respectively, the CR proton and electron ionisation rates can be expressed as:
\begin{equation}
\begin{split}
    \zeta_\mathrm{p} &= \int_{E_\mathrm{min}}^{E_\mathrm{max}}v_\mathrm{p}f_\mathrm{p}(E_\mathrm{p})\sigma_\mathrm{p}^{ion}(E_\mathrm{p})(1+\phi_\mathrm{p}(E_\mathrm{p}))\,dE_\mathrm{p} \\
    &+ \int_{0}^{E_\mathrm{max}}v_\mathrm{p}f_\mathrm{p}(E_\mathrm{p})\sigma_\mathrm{p}^{e.c.}(E_\mathrm{p})\,dE_\mathrm{p} \\
    \zeta_\mathrm{e} &= \int_{E_\mathrm{min}}^{E_\mathrm{max}}v_\mathrm{e}f_\mathrm{e}(E_\mathrm{e})\sigma_\mathrm{e}^{ion}(E_\mathrm{e})(1+\phi_\mathrm{e}(E_\mathrm{e}))\,dE_\mathrm{e} 
\end{split}
\label{eq:zeta}
,\end{equation} 
where $v_\mathrm{i}$ is the particle velocity, %$f_\mathrm{i}$ the CR density, 
and $\phi_\mathrm{i}$ the average number of secondary ionisations per primary ionisation, modelled as in \cite{kra15}.
Available parameterisations of the proton impact ionisation cross-section, $\sigma_\mathrm{p}^{ion}$, the electron capture cross-section, $\sigma_\mathrm{p}^{e.c.}$, and the electron impact ionisation cross-sections, $\sigma_\mathrm{e}^{ion}$, have been reviewed by \cite{pad09} and later improved by \citet{kra15} and \citet{gab22}. 
The impact of CR nuclei heavier than protons should also be accounted for when computing the CR ionisation rate.
This was done by introducing an enhancement factor, $\eta\approx1.5$, to make the CR ionisation rate of nuclei (including protons) $\zeta_\mathrm{n}=\eta\zeta_\mathrm{p}$ \citep{pad09}. In the following, the term proton ionisation rate will refer to the nuclear-enhanced value.

The electron ionisation cross-section peaks in the sub-kilo-electronvolt energy range, and any electron of energy larger than the ionisation potential of $\Htwo$, $I = 15.426~\mathrm{eV}$ will contribute to ionising the ambient gas. Hence, we considered CR electrons of energy larger than $E_\mathrm{min}=I$.

On the other hand, for ionisation due to CR protons, electron capture dominates over proton impact for particle energies below a few tens of kilo-electronvolts (see e.g. Fig.~11 in \citealt{gab22}).
This means that a CR proton cooling below that energy will convert into a fast neutral hydrogen atom and, as a consequence, will become unable to further ionise the gas \citep[e.g.][]{cha16}.
For this reason, for protons, we set $E_\mathrm{min} = 45~\mathrm{keV}$.

As a first step, the CR proton and electron injection spectra inferred from $\gamma$-ray and radio data in the previous section have been extrapolated down to lower energies and propagated using the transport equation solver to obtain steady-state spectra. Then, the CR proton and electron ionisation rates averaged over the entire CMZ volume were computed to find $1.9\times10^{-17}~\mathrm{s}^{-1}$ and $3.0\times10^{-18}~\mathrm{s}^{-1}$, respectively.
These values are much (orders of magnitude) smaller than the measured ones, meaning that an extra component of CRs must be present in order to explain the observations.

The extra component (enhancement) was added to the CR injection spectra derived above in the following way:
\begin{equation}
        Q_\mathrm{p,p}(p)= 
\begin{dcases}
    Q_\mathrm{enh,p}\left(\frac{p}{p_\mathrm{enh,p}}\right)^{-\delta_\mathrm{enh,p}} ,& \text{if } p\leq p_\mathrm{enh,p}\\
     Q_\mathrm{enh,p}\left(\frac{p}{p_\mathrm{enh,p}}\right)^{-\delta_\mathrm{p}} ,& \text{otherwise}
\end{dcases}
\end{equation} 
for protons and
\begin{equation}
        Q_\mathrm{p,e}(p)= 
\begin{dcases}
    Q_\mathrm{enh,e}\left(\frac{p}{p_\mathrm{enh,e}}\right)^{-\delta_\mathrm{enh,e}} ,& \text{if } p\leq p_\mathrm{enh,e}\\
     Q_\mathrm{enh,e}\left(\frac{p}{p_\mathrm{enh,e}}\right)^{-\delta_\mathrm{e,1}} ,& \text{if } p_\mathrm{enh,e}\leq p \leq p_{*} \\
     Q_\mathrm{enh,e}\left(\frac{p_*}{p_\mathrm{enh,e}}\right)^{\delta_\mathrm{e,2}-\delta_\mathrm{e,1}}\left(\frac{p}{p_\mathrm{enh,e}}\right)^{-\delta_\mathrm{e,2}} ,& \text{otherwise}
\end{dcases}
\end{equation} 
for electrons.

\begin{figure}[htbp]
  \centering
  \begin{subfigure}{\hsize}
    \centering
    \includegraphics[width=\hsize]{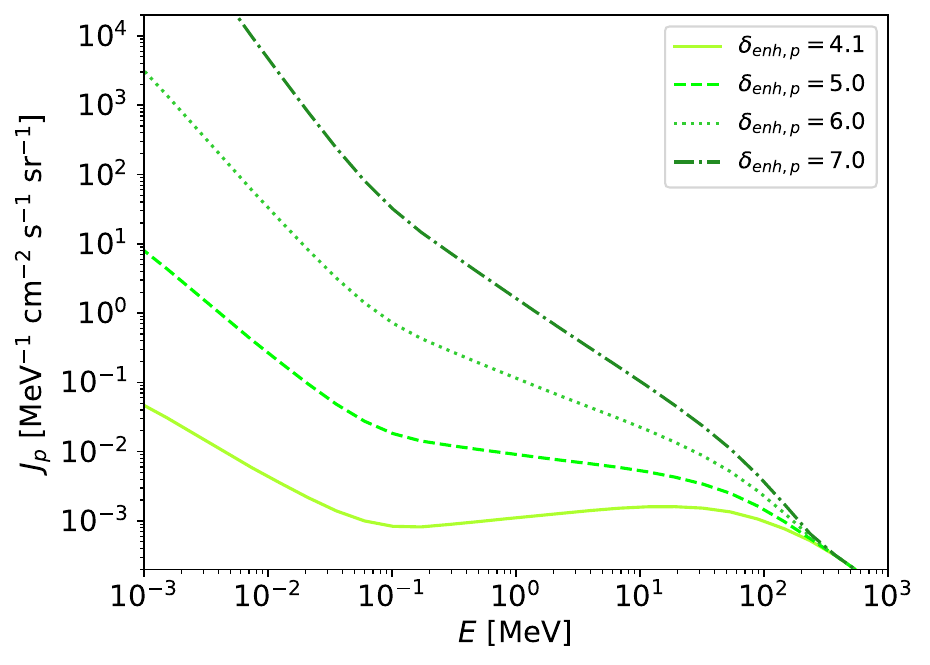}
    \label{specp}
  \end{subfigure}
  \hfill
  \begin{subfigure}{\hsize}
    \centering
    \includegraphics[width=\hsize]{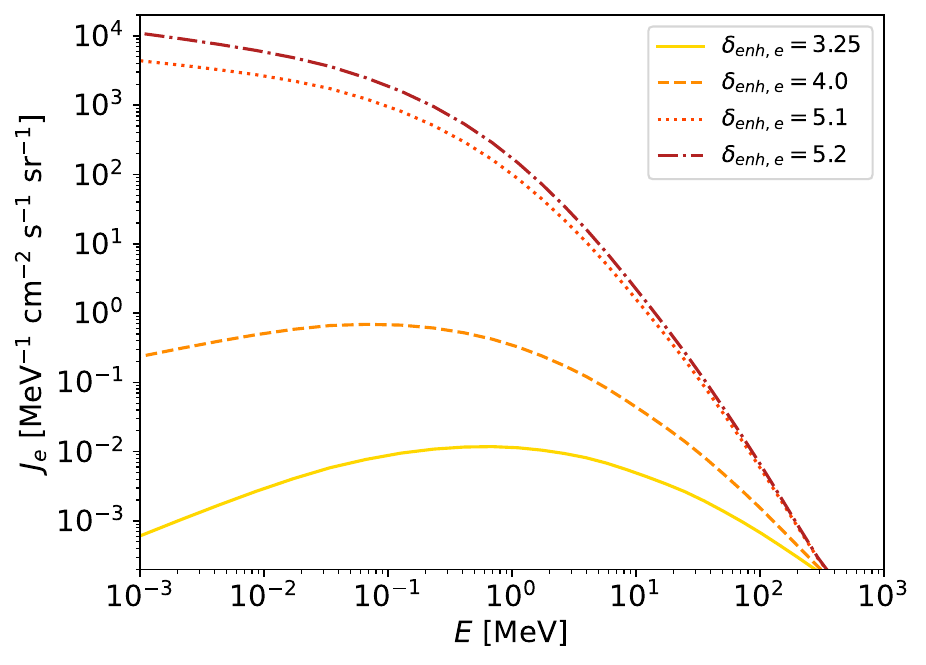}
    \label{spece}
  \end{subfigure}
  \caption{Steady-state CR intensities for protons (top) and electrons (bottom) averaged over the CMZ volume. Curves refer to different enhancements applied to the CR injection spectrum, as is indicated by the parameter $\delta_\mathrm{enh,i}$.}
  \label{spec}
\end{figure}

For particle momenta below $p_{enh, i}$, CR spectra were enhanced with respect to those adopted so far in this paper.
For protons, we chose $p_\mathrm{enh,p}c\approx780~\mathrm{MeV}$. This momentum corresponds to a kinetic energy of $E_\mathrm{p}=280~\mathrm{MeV}$, which is the threshold for pion production. Adding an enhancement below this momentum will ensure that no $\gamma$-rays are produced by the additional component, and therefore the fits to $\gamma$-ray data performed above will be unaffected.
For electrons, the transition momentum was chosen to be equal to $p_\mathrm{enh,e}c\approx450~\mathrm{MeV}$, which ensures that the fit to radio data obtained above is unaffected.
This can be seen in Fig.~\ref{fig:radiofs}, in which the synchrotron radio emission from enhanced electron spectra is shown together with radio data.

The steady-state CR proton and electron intensities averaged over the entire CMZ volume are shown in Fig.~\ref{spec} for various values of the parameter $\delta_\mathrm{enh,i}$.

\begin{figure}[htbp]
  \centering
  \begin{subfigure}{\hsize}
    \centering
    \includegraphics[width=\hsize]{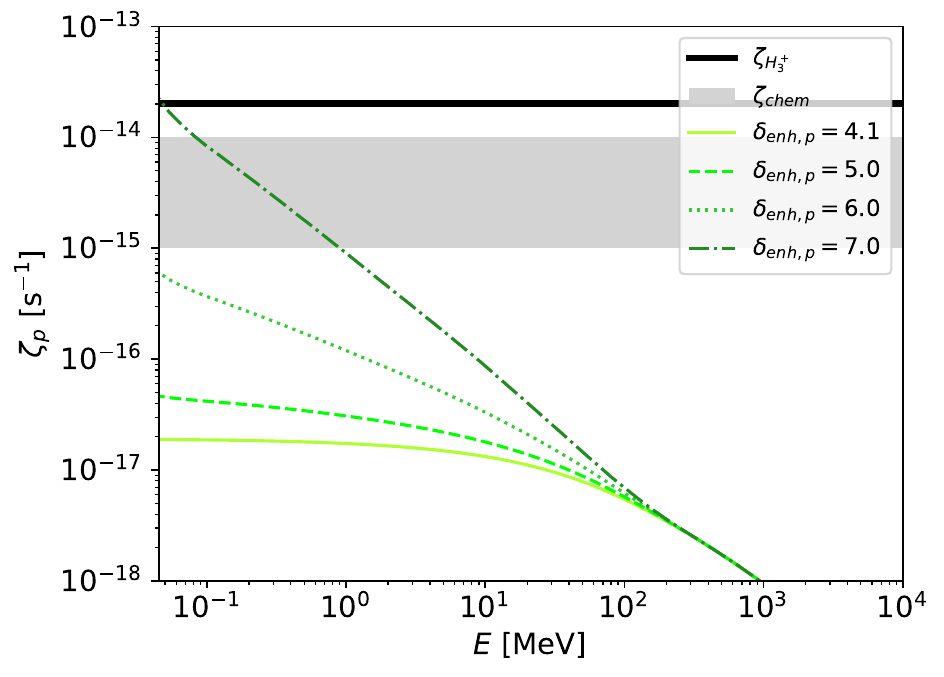}
    \label{ionp}
  \end{subfigure}
  \hfill
  \begin{subfigure}{\hsize}
    \centering
    \includegraphics[width=\hsize]{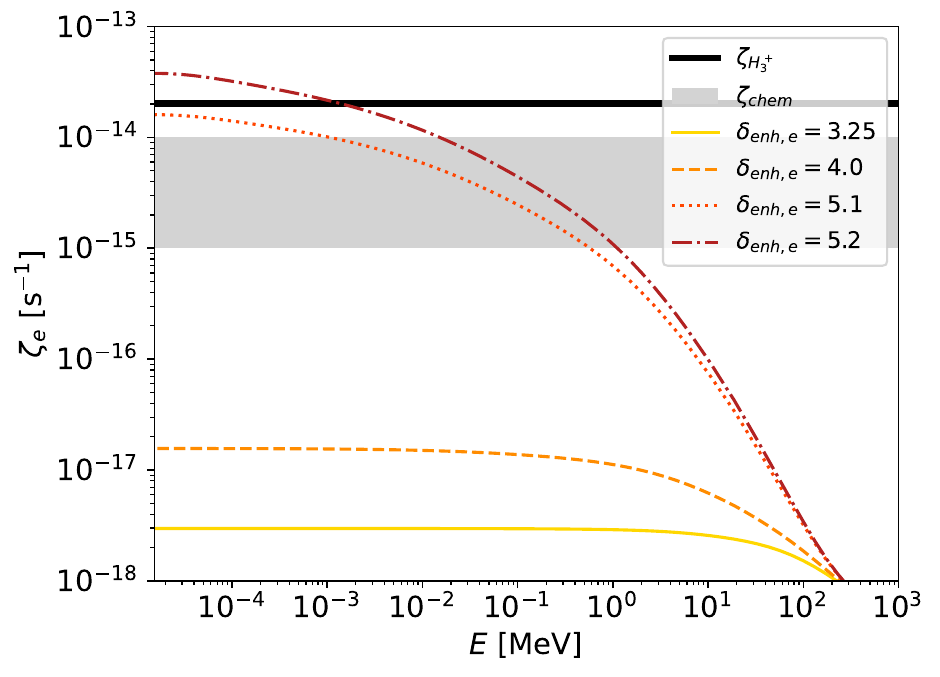}
    \label{ione}
  \end{subfigure}
  \caption{CR proton (top) and electron (bottom) ionisation rates as a function of the minimum particle energy considered for the CR spectra shown in Fig. \ref{spec}.
  Values were averaged over the CMZ volume and compared to the estimated values of ionisation rates derived by different observational methods (horizontal black line and shaded region).}
  \label{ion}
\end{figure}

The proton and electron ionisation rates computed using these volume-averaged spectra are shown in Fig. \ref{ion} as a function of different minimum ionising particle energies, $E_\mathrm{min}$ (this is the minimum energy that appears in Eq.~\ref{eq:zeta}). 
For comparison, we have also plotted as a horizontal thick black line the ionisation rate, $\zeta_\mathrm{H_3^+}$, in the GC region obtained from $\rm H_3^+$ observations, obtained by averaging the measurements from \cite{lep16, oka19}. We have also added a grey-shaded region, which gives the range of values for the more conservative ionisation rates, $\zeta_\mathrm{chem}$, inferred from measurements of the abundances of other chemical species \citep{riv22, san24}.

We see from Fig.~\ref{ion} that in order to reproduce the ionisation rate measurements, one needs an extremely steep additional CR component (slopes larger than $\delta_\mathrm{enh,p} \sim 7$ and $\delta_\mathrm{enh,e} \sim 5$ for CR protons and electrons, respectively) extending down to very low particle energies ($E_\mathrm{min} \approx$~1~keV $-$ 1~MeV for both CR protons and electrons).
Very steep spectra are needed because the energy loss time of CR particles is very short at sub-giga-electronvolt energies (see Fig.~\ref{times}), and therefore this has to be compensated for by the injection of a very large number of low-energy particles \citep[for a discussion of this issue see][]{rec19}.
This raises the question of what energy budget is required to maintain a population of particles capable of reproducing the measured ionisation rate in the CMZ.

To answer this question, we computed two quantities: the power, $W_\mathrm{CR}$, needed to maintain the CR population at a given level in the CMZ and the CR energy density averaged over the entire CMZ volume, $n_\mathrm{CR}$.
These quantities were computed for CR spectra able to reproduce the measurements of the ionisation rates in the CMZ and are reported in Tabs.~\ref{tab:ergp} and \ref{tab:erge} for CR protons and electrons, respectively. Here, we note $E_\mathrm{min}$, the minimum ionising energy needed for the computed ionisation rate to exceed the measured values.

\begin{table}
\caption{Power requirements and energy density for proton spectra as a function of the corresponding proton CR ionisation rate in the CMZ}             
\label{pres}      
\centering                          
\begin{tabular}{c c c c c}        
\hline\hline               
$\zeta_\mathrm{p}(\mathrm{s}^{-1})$ & $\delta_\mathrm{enh,p}$ & $E_\mathrm{min}(\mathrm{MeV})$ & $W_\mathrm{CR} (\mathrm{erg}~\mathrm{s}^{-1})$ & $n_\mathrm{CR} (\mathrm{eV}\,\mathrm{cm}^{-3})$ \\    
\hline                     
   $10^{-15}$ & 7.0 & $0.9$ & $3.7~10^{39}$ & $0.48$ \\
   $2~10^{-14}$ & 7.0 & $4.5~10^{-2}$ & $6.8~10^{40}$ & $0.52$ \\
\hline                          
\end{tabular}
\label{tab:ergp}
\end{table}

The powers obtained in this way can be compared with the total CR power of the Milky Way, which can be estimated using the large amount of available CR data.
This has been done, among many others, by \cite{str10}, who found values in the range of $W_\mathrm{MW}^p \sim 6.0 - 7.4 \times 10^{40}$~erg~s$^{-1}$ and $W_\mathrm{MW}^e \sim 1.0 - 1.7 \times 10^{39}$~erg~s$^{-1}$ for CR protons and electrons, respectively.
The comparisons with these figures show that the powers reported in Tabs.~\ref{tab:ergp} and \ref{tab:erge} are extremely large.

\begin{table}
\caption{Power requirements and energy density for electron spectra as a function of the corresponding electron CR ionisation rate in the CMZ}             
\label{eres}      
\centering                     
\begin{tabular}{c c c c c}     
\hline\hline           
$\zeta_\mathrm{e}(\mathrm{s}^{-1})$ & $\delta_\mathrm{enh,e}$ & $E_\mathrm{min}(\mathrm{MeV})$ & $W_\mathrm{CR} (\mathrm{erg}~\mathrm{s}^{-1})$ & $n_\mathrm{CR} (\mathrm{eV}\,\mathrm{cm}^{-3})$ \\    
\hline                       
   $10^{-15}$ & $5.1$ & $0.6$ & $2.0~10^{39}$ & $0.36$ \\ 
   $10^{-15}$ & $5.2$ & $1.1$ & $2.1~10^{39}$ & $0.45$ \\
   $2~10^{-14}$ & $5.2$ & $1.5~10^{-3}$ & $2.4~10^{40}$ & $0.58$ \\
\hline                       
\end{tabular}
\label{tab:erge}
\end{table}

In order to reproduce the values of the ionisation rate measured in the CMZ, CR electrons should be injected by the source located in the GC at a rate that is larger than the total injection rate of CR electrons in the entire Milky Way. For CR protons, the required power has to be comparable to the total CR proton power of the Milky Way in order to reproduce an ionisation rate of the order of $10^{-14}$~s$^{-1}$, and a few percent of it to reproduce an ionisation rate of the order of $10^{-15}$~s$^{-1}$.
Moreover, this huge amount of energy should have as a unique manifestation the enhancement in the gas ionisation rate in the CMZ region.

One can also compare the CR powers obtained above with characteristic powers connected to the SMBH at the centre of the Milky Way.
It has been estimated that the mechanical power needed to inflate the giant ($\gtrsim 10$~kpc) bubbles discovered by Fermi \citep{su10} and eROSITA \citep{pre20} is of the order of $P_\mathrm{b} \approx 10^{41}$~erg~s$^{-1}$, and that the injection of energy should last a few tens of millions of years \citep{sar24}.
This is of the same order as the energy required to maintain a population of CR protons capable of ionising the CMZ gas at a rate of $\approx 10^{-14}$~s$^{-1}$ and about four times the energy needed in CR electrons to reach the same rate (see last row of both Tab.~\ref{tab:ergp} and \ref{tab:erge}).
It follows that it is quite challenging to explain the large ionisation rates with CRs, while lower values ($\approx 10^{-15}$~s$^{-1}$) might still be acceptable.

However, another observable that should be reproduced is the fact that roughly the same ionisation rate has been obtained by analysing H$_3^+$ lines from several lines of sight across the CMZ (at least up to 100 pc radius, \citealt{ind15}).
The ionisation rate measured in this way does not show any clear spatial variation and is indeed compatible with a constant value across the CMZ.
This trend cannot be reproduced by our model, as is shown in Fig.~\ref{ionR}.
There, we plot the CR ionisation rate averaged along the lines of sight lying along the Galactic plane at different distances, $R$, from the GC.
The predicted trend for CR protons (top) and electrons (bottom panel) clearly does not explain observations.

We note here that the mass of the CMZ that has been chosen to derive the values reported in Tabs.~\ref{tab:ergp} and \ref{tab:erge}, though widely used in the literature, is the maximum amongst the various measurements found in the literature \citep[which is $2-6\times10^7~\mathrm{M}_\odot$, as reported in][]{dah98}. 
We repeated our entire study for the lower limit of $M_\mathrm{CMZ}=2\times10^{7}~\mathrm{M}_\odot$ and found that the power needed to sustain ionisation rates at $\approx 10^{-14}$~s$^{-1}$ is also in this case of the order of $\gtrsim10^{40}~\mathrm{erg}\,\mathrm{s}^{-1}$ for protons and $\gtrsim10^{39}~\mathrm{erg}\,\mathrm{s}^{-1}$ for electrons.
This is because a lower mass (and therefore gas density) would require a higher density of high-energy CRs to explain the $\gamma$-ray and radio emissions, but at the same time would reduce the effects of ionisation losses.
The two effects then compensate.

\begin{figure}[htbp]
\centering
   \includegraphics[width=0.8\hsize]{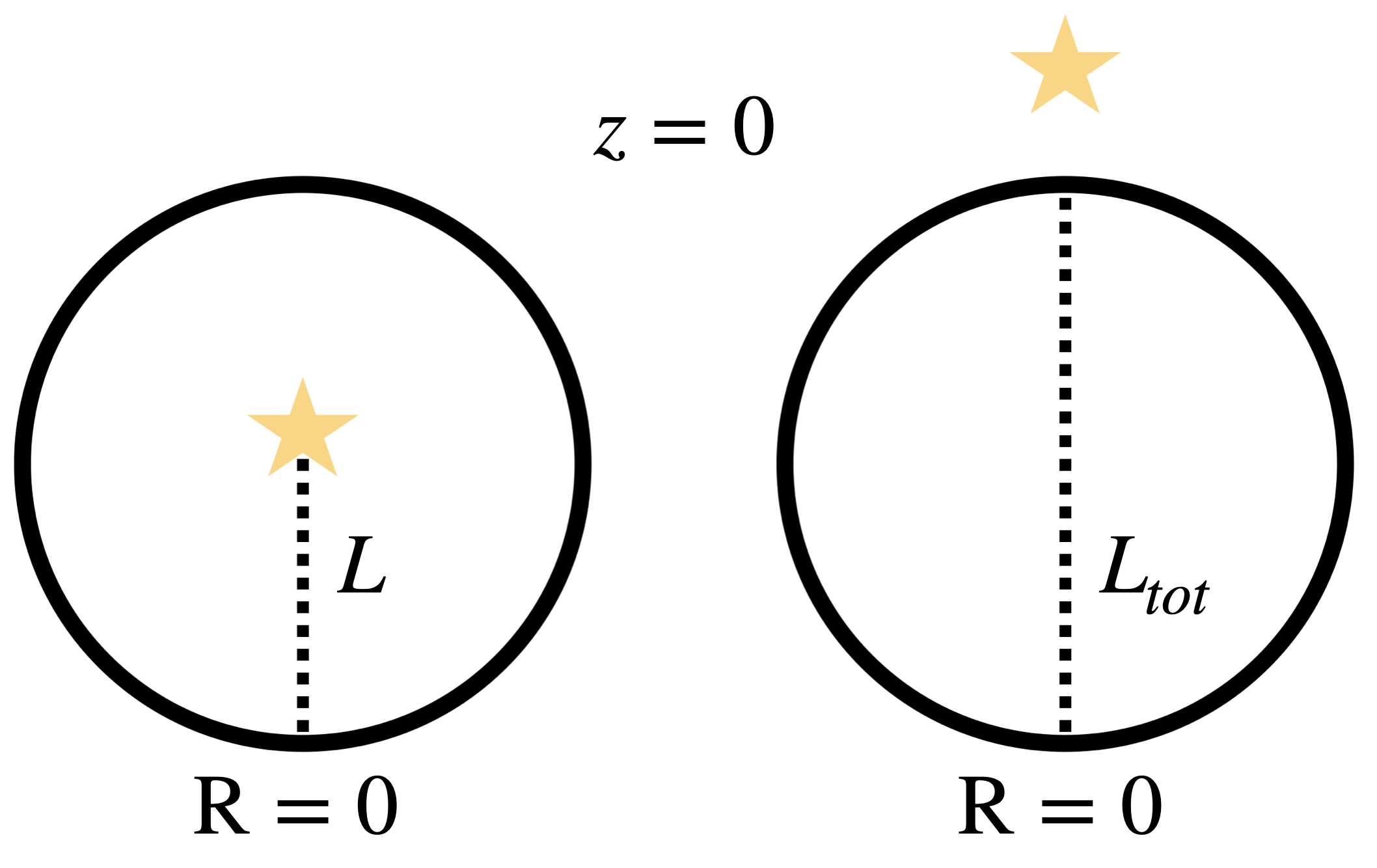}
     \caption{Position of the star used to measure the column density of H$_3^+$ with respect to the CMZ. $L$ is the distance from the front of the CMZ to the position of the star. $L_\mathrm{tot}=2R_\mathrm{CMZ}$ is the longest distance through the CMZ in any line of sight.}
     \label{fig:star}
\end{figure}

\begin{figure}[htbp]
  \centering
  \begin{subfigure}{\hsize}
    \centering
    \includegraphics[width=\hsize]{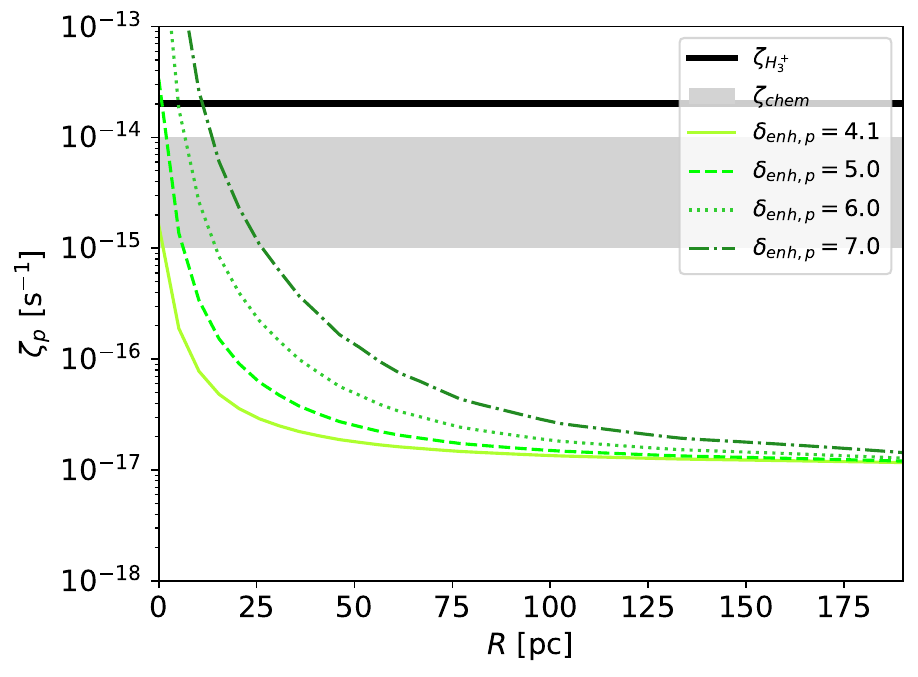}
    \label{ionpr}
  \end{subfigure}
  \hfill
  \begin{subfigure}{\hsize}
    \centering
    \includegraphics[width=\hsize]{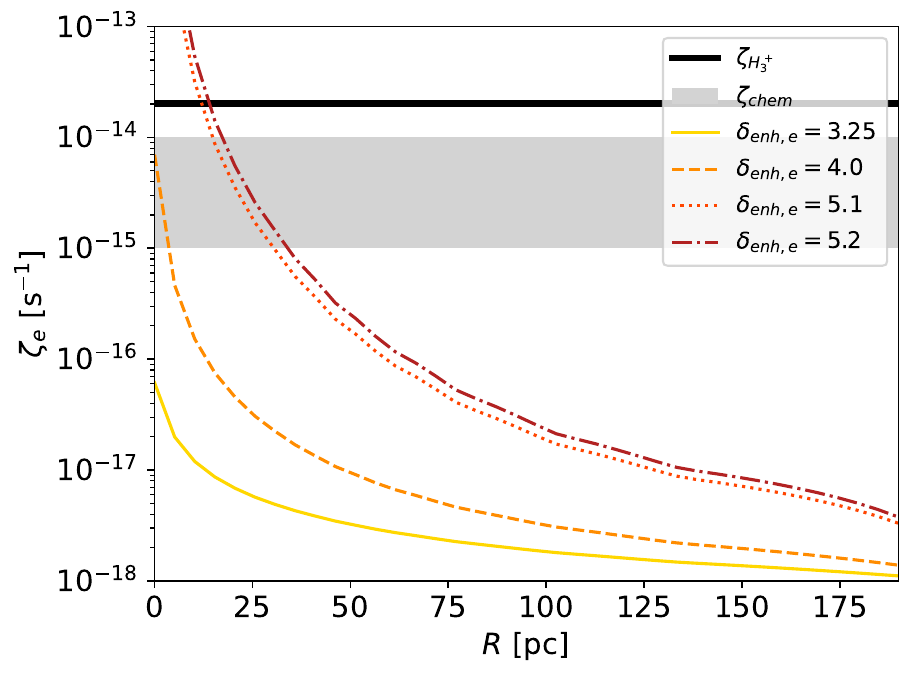}
    \label{ioner}
  \end{subfigure}
  \caption{CR proton (top) and electron (bottom) ionisation rates averaged over lines of sight at different projected distances, $R$, from the GC. The horizontal lines and shaded regions and the values of $\delta_\mathrm{enh,i}$ are as in Fig.~\ref{ion}.}
  \label{ionR}
\end{figure}

Before concluding, we should notice that the measurements of H$_3^+$ column densities rely on the presence of a background star.
However, the distance to these stars is a major source of uncertainty in the measurements.
In fact, the results we presented so far assumed implicitly that the stars are located behind the CMZ (as is illustrated in the right panel of Fig.~\ref{fig:star}).
If, instead, the star is located within the CMZ (as in the left panel of Fig.~\ref{fig:star}), the average of the predicted CR ionisation rate should be performed over the length, $L$.

\begin{figure}[htbp]
  \centering
  \begin{subfigure}{1.03\hsize}
    \centering
    \includegraphics[width=\hsize]{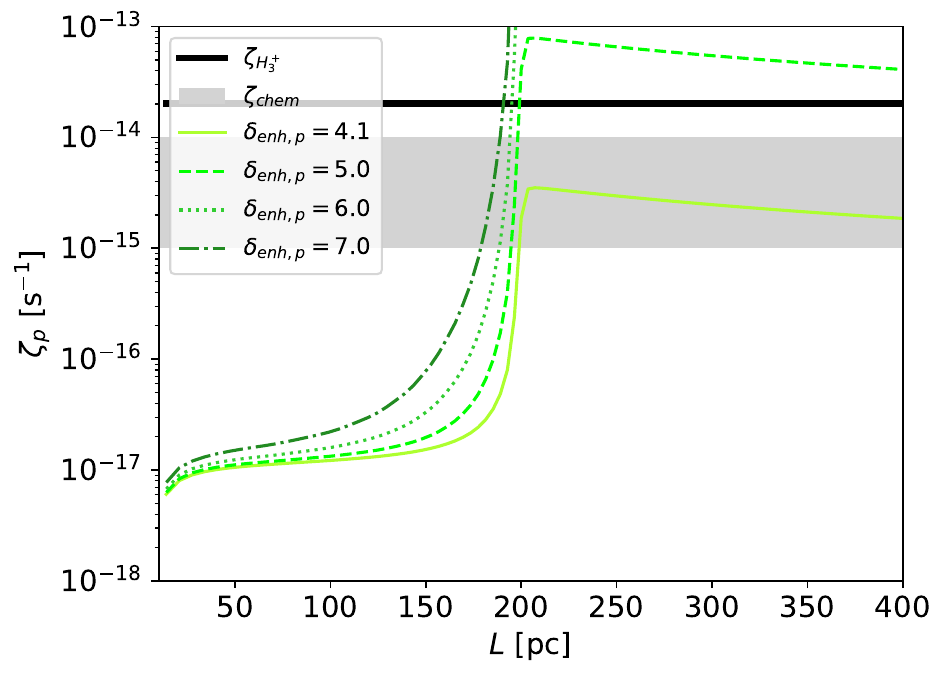}
    \label{ionpl}
  \end{subfigure}
  \hfill
  \begin{subfigure}{1.03\hsize}
    \centering
    \includegraphics[width=\hsize]{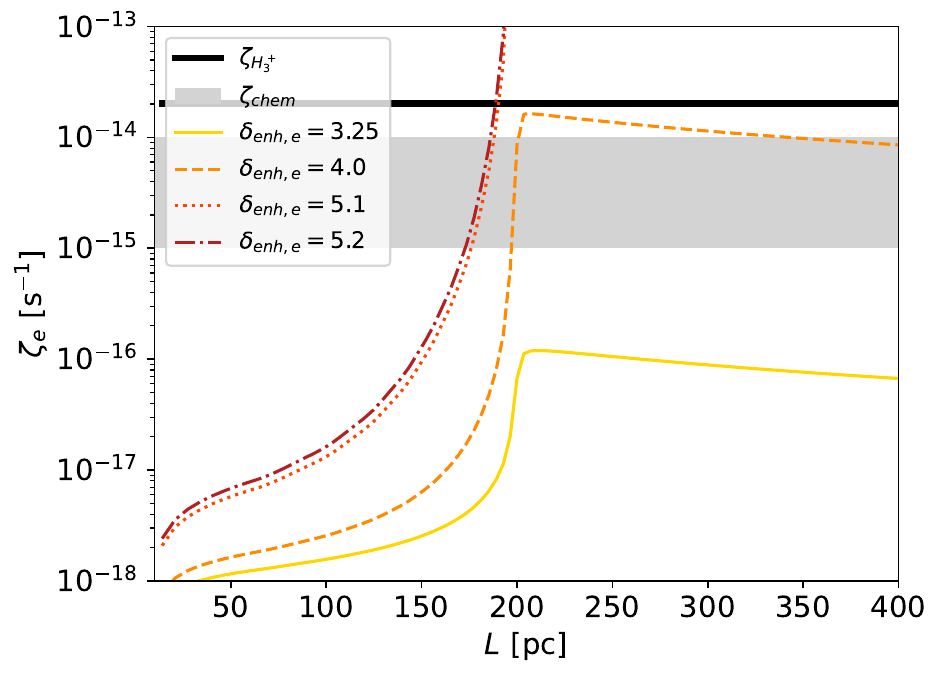}
    \label{ionel}
  \end{subfigure}
  \caption{CR proton (top) and electron (bottom) ionisation rates averaged over different lengths of lines of sight. The horizontal lines and shaded regions and the values of $\delta_\mathrm{enh,i}$ are as in Fig.~\ref{ion}.}
  \label{ionL}
\end{figure}

Figure~\ref{ionL} shows the CR ionisation rates averaged from the edge of the CMZ to a depth, $L$, for lines of sight passing through the point $z = 0$ and $R = 0$.
In such a plot, the ionisation rate averaged over the entire line of sight is represented by the rightmost value of the curves.
The figure shows that if the star is located closer to us than the GC, then one would expect to measure much lower ionisation rates, while only a very slight enhancement of the average ionisation rate is expected for locations of $L \gtrsim R_\mathrm{CMZ}$.
This makes it even more difficult (if not impossible) to explain the large ionisation rates derived from observations.
The plot refers to a line of sight passing through the centre of the CMZ ($R = 0$ and $z = 0$), and therefore corresponds to the maximum possible values of the ionisation rate (as is illustrated by Fig.~\ref{ionR}).

\section{Additional constraints on mega-electronvolt particles}
\label{sec:add}

\subsection{Mega-electronvolt $\gamma$-rays}

The CRs in the mega-electronvolt and sub-mega-electronvolt domains contribute the most to the ionisation rate.
Electrons of this energy also emit non-thermal bremsstrahlung photons in the hard-X/soft-$\gamma$-ray bands.
At such energies, the inner Galaxy has been observed by the SPectrometer on INTEGRAL (SPI) and by the imaging Compton telescope COMPTEL.
In particular, \citet{bou08} presented a list of sources detected by SPI at photon energies in the range spanning from 25 keV to 0.6 MeV.
Among those, three are located in the CMZ region: 1E 1740.7$-$2942 ($l=359.12^\circ, b=-0.1^\circ$), SLX 1744$-$299 ($l=359.28^\circ, b=-0.9^\circ$), and IGR J17475$-$2822 ($l=0.61^\circ,b=-0.1^\circ$).
The fluxes of these sources in several distinct energy bands can be found in Tab.1 in \citet{bou08}.
From that table, we also searched for the weakest detected source in the inner Galaxy, and we took that flux as a proxy for the instrument sensitivity.

We predict a flux for the CMZ (for the case $\delta_\mathrm{e}=5.2$) equal to $4\times 10^{-2}~\mathrm{keV}~\mathrm{cm}^{-2}~\mathrm{s}^{-2}$ in the $25-50~\mathrm{keV}$ band, $5\times 10^{-2}~\mathrm{keV}~\mathrm{cm}^{-2}~\mathrm{s}^{-2}$ in the $50-100~\mathrm{keV}$ band, $6\times 10^{-2}~\mathrm{keV}~\mathrm{cm}^{-2}~\mathrm{s}^{-2}$ in the $100-200~\mathrm{keV}$ band, and $8\times 10^{-2}~\mathrm{keV}~\mathrm{cm}^{-2}~\mathrm{s}^{-1}$ in the $200-600~\mathrm{keV}$ band.
These fluxes are larger than the instrument sensitivity for point sources, and even larger than some of the fluxes of the three sources detected in the CMZ and listed above.
In particular, the flux detected by SPI from IGR J17475$-$2822, associated with the Sgr~B2 MCs, is known to decline with time \citep{ter10}. This is not compatible with a CR-related origin of the emission.

To conclude, if the spectra predicted by our model were the reality, the CMZ would have been detected by SPI as a steady source of mega-electronvolt $\gamma$-rays with a centroid around $l=0^\circ$ and $b=0^\circ$. Hence, the conclusion stands that mega-electronvolt $\gamma$-ray detection does not support the CR ionisation model.

%\begin{figure}[htbp]
%   \centering
%   \includegraphics[width=\hsize]{figs/mevg.pdf}
%      \caption{Bremsstrahlung $\gamma$-rays from electron spectra capable of giving ionisation rate above $10^{-15}~\mathrm{s}^{-1}$ compared to SPI and COMPTEL data from the inner galaxy $\vert{l}\vert < 30^\circ$ and  $\vert{b}\vert < 15^\circ$ from \cite{bou11}.}
%        \label{fig:mevgam}
%   \end{figure}

\subsection{Iron K$\alpha$ line}

The X-ray emission line at $6.4~\mathrm{keV}$ results from the ionisation of iron atoms by the ejection of a K-shell electron. The diffuse X-ray emission from the CMZ has a variable and a constant component. 
The variable component is believed to be the reflection by the MCs of a past X-ray outburst from the SMBH, which is why it is decreasing over time \citep{ter18}. 
A constant component is expected from the supposed constant density of CRs ionising the region \citep{cap12}. 
Since the observations of the CMZ over several years show a decreasing flux, the most recent measurement of the Fe K$\alpha$ line flux is only an upper limit for the contribution from the CRs. 
The most recent observation of the inner $300~\mathrm{pc}$ was conducted by XMM-Newton in 2012. 
Other more recent surveys of specific regions exist \citep{kuz22}. 
The XMM-Newton survey has revealed that the $6.4~\mathrm{keV}$ line flux in 2012 was $1.507\pm0.009\times10^{-3}~\mathrm{ph}\,\mathrm{cm}^{-2}\,\mathrm{s}^{-1}$ from a $19\times112~\mathrm{arcmin}^2$ area \citep{ter18}. As this region is very similar to the CMZ region, we may compare the average surface brightness from our model with the one measured by XMM-Newton: the corresponding average surface brightness is $7.36\pm0.04\times10^{-7}~\mathrm{ph}\,\mathrm{cm}^{-2}\,\mathrm{s}^{-1}\,\mathrm{arcmin}^{-1}$.

The intensity of the Fe K$\alpha$ line resulting from CR interactions can be expressed as:
\begin{equation}
    I_\mathrm{i}^{K\alpha} = \frac{M_\mathrm{CMZ}}{4\pi{D_\mathrm{GC}^2}{m_\mathrm{avg}}}\int_{E_\mathrm{min}}^{E_\mathrm{max}}\eta_{Fe}\sigma_\mathrm{i}^{K\alpha}(E_\mathrm{i})v_\mathrm{i}f_\mathrm{i}(E_\mathrm{i})\,dE_\mathrm{i}
    ,\end{equation}
where $i$ represents the species of the CR particle, $\sigma_{K\alpha}^i$ is the K-shell ionisation cross-section by CR species i \citep{tat12} considering the solar abundance of iron ($\eta_{Fe}=3 \times 10^{-5}$), and $m_\mathrm{avg}=1.4m_\mathrm{H}$ is the average particle mass. 

We computed the Fe K$\alpha$ average surface brightness from the CMZ using the proton and electron spectra that give ionisation rates above $10^{-15}~\mathrm{s}^{-1}$. 
The values are given in Tables \ref{pFe} and \ref{eFe}. 
The iron abundance in the CMZ is expected to be higher than that of the solar neighbourhood. 
Hence, the values of the Fe K$\alpha$ line emissions are a lower limit for what is expected from the CR spectra.

\begin{table}[h]
\caption{Fe K$\alpha$ average surface brightness from CR protons}           
\label{pFe}     
\centering                         
\begin{tabular}{c c c c}        
\hline\hline                
$\zeta_\mathrm{p}(\mathrm{s}^{-1})$ & $\delta_\mathrm{enh,p}$ & $E_\mathrm{min}(\mathrm{MeV})$  & $B_\mathrm{p}^{K\alpha}(\mathrm{ph}\,\mathrm{cm}^{-2}\,\mathrm{s}^{-1}\,\mathrm{arcmin}^{-2})$ \\    
\hline                      
   $10^{-15}$ & $7.0$ & $0.9$  & $5.8~10^{-10}$ \\
   $2~10^{-14}$ & $7.0$ & $4.5~10^{-2}$ & $5.8~10^{-10}$ \\
\hline                         
\end{tabular}
\end{table}

\begin{table}[h]
\caption{Fe K$\alpha$ average surface brightness from CR electrons}            
\label{eFe}    
\centering                        
\begin{tabular}{c c c c}        
\hline\hline                
$\zeta_\mathrm{e}(\mathrm{s}^{-1})$ & $\delta_\mathrm{enh,e}$ & $E_\mathrm{min}(\mathrm{MeV})$ & $B_\mathrm{e}^{K\alpha}(\mathrm{ph}\,\mathrm{cm}^{-2}\,\mathrm{s}^{-1}\,\mathrm{arcmin}^{-2})$ \\   
\hline      
    $10^{-15}$ & $5.1$ & $0.6$ & $7.8~10^{-9}$  \\
   $10^{-15}$ & $5.2$ & $1.1$ & $8.3~10^{-9}$  \\
   $2~10^{-14}$ & $5.2$ & $1.5~10^{-3}$ & $5.1~10^{-8}$ \\
\hline                         
\end{tabular}
\end{table}

The expected Fe K$\alpha$ surface brightness over the CMZ resulting from CR ionisation is lower by a few orders of magnitude than the observed rate. 
This estimate may increase if the correct iron abundance is used instead of solar abundance, but the enhancement would be, at most, a few factors, and therefore not sufficient to enhance our prediction of the observed values. 
Hence, Fe K$\alpha$ line observations cannot help to constrain the spectrum of low-energy CRs in the CMZ. Indeed, this agrees with what was claimed by \citet{dog13}, in which it was argued that the contribution to the $6.4~\mathrm{keV}$ line from low-energy CRs responsible for the high ionisation rates must be negligible.

%\section{Discussion}

\section{Discussion and conclusion}
\label{sec:ccl}

The main conclusion of this paper is that it is extremely unlikely that CRs in the CMZ are the agents responsible for the large ionisation rates derived from a number of observations. Motivated by the results coming from $\gamma$-ray observations of the CMZ, we investigated a scenario in which a powerful accelerator of CR is located in the centre of the Galaxy.
The accelerator is assumed to continuously inject CRs in the surrounding medium.
Once injected, CRs diffuse away from the GC and fill the CMZ.

Fitting the $\gamma$-ray and radio emission from the CMZ allowed us to constrain the high-energy spectra of CR particles there.
We found that a simple power-law extrapolation of such spectra at low energies fails by orders of magnitude to reproduce the measured values of the ionisation rate of molecular hydrogen. 

We then added an additional, steep, low-energy CR component in an attempt to explain the large values of the ionisation rate without violating any other observational constraint.
We showed that the injection power of energetic particles required to explain ionisation rates at the level of $\approx 10^{-14}$~s$^{-1}$ is exceedingly large: $\sim7\times10^{40}~\mathrm{erg}\,\mathrm{s}^{-1}$ for protons and $\sim2\times10^{40}~\mathrm{erg}\,\mathrm{s}^{-1}$ for electrons. This is comparable to the total power, $P_\mathrm{b}$, required to inflate the giant eROSITA bubbles.
More conservative estimates of the ionisation rate ($\approx 10^{-15}$~s$^{-1}$) could be explained if a few percent of the power, $P_\mathrm{b}$, could be somehow converted into CRs.

In fact, this unrealistically large power estimation for low-energy CRs is obtained when using the most conservative values for the transport parameters. We have chosen the diffusion coefficient, the Galactic wind velocity, and the Alfv\'en speed to maximise the duration for which CRs stay within the CMZ. 
Hence, we argue that using different values for all these parameters would further increase the power required.

However, the CR energy budget is not the only issue.
The CRs escaping from an accelerator located at the GC would generate an ionisation rate that declines quite steeply as the distance from the GC increases.
This is not observed in the data, which show a roughly constant value of the ionisation rate throughout the CMZ.
Uncertainties in the exact location of the stars used make it even more difficult to fit our model results to the data. In fact, these high ionisation rates are also expected in the cores of dense MCs, and \citet{yan23} show that CRs below $\lesssim \mathrm{GeV}$ are prevented from entering such MCs.

We note that the CR energy densities found in our model (see last column of Tabs.~\ref{pres} and \ref{eres}) are quite modest. 
This was also pointed out in a previous study by \citet{dog15}.
However, we have stressed here that the power needed to maintain these modest energy densities of CRs is extremely large, mainly due to the very short energy loss time of sub-relativistic particles (see Fig.~\ref{fig:losstimes} and {times}), and that this makes a CR origin of the large ionisation rates very unlikely.

A different scenario could be envisaged whereby an impulsive rather than continuous source of CRs is present.
In that case, however, the large ionisation rates would be a transient phenomenon, and this would add an additional parameter (i.e. the time since the impulsive injection) to be fine-tuned.
An inspection of Fig.~\ref{times} can help in characterising what could be the optimal set-up for this scenario.
The duration of the enhanced ionisation rates would be of the order of $\Delta t \lesssim 10^5$~yr, which is the characteristic diffusion time of $\approx 100$~MeV CRs across the CMZ.
Such particles would lose their energy in a time comparable to the diffusion time across the CMZ.
A naive but probably not too inaccurate estimate of the total required energy in this scenario is given by the product ${\cal E}_\mathrm{tot} \approx W_\mathrm{CR} \times \Delta t$, where for $W_\mathrm{CR}$ one can adopt the values reported in Tab.~\ref{eres} and \ref{pres}.
This would give ${\cal E}_\mathrm{tot} \approx 3 \times 10^{52} (W_\mathrm{CR}/10^{40}~{\rm erg/s}) (\Delta t/10^5 {\rm yr})$~erg.
Remarkably, this is of the same order as the thermal energy of the X-ray chimneys ($\sim 4 \times 10^{52}$~erg), which are characterised by a sound crossing time equal to $\sim 3 \times 10^5$~yr, which is comparable to $\Delta t$ \citep{pon19}.
The chimneys are the exhaust channels through which mass and energy ejected from the SMBH in the GC are channelled out of the Galactic disc.
The similarity between their total energy and that needed in low-energy CRs suggests that a scenario based on an impulsive injector of particles would also face severe problems based on global energetic constraints.

Finally, the rough spatial uniformity of the values of the ionisation rate could be explained if many CR sources distributed across the entire CMZ inject energetic particles.
In this case, the energy problem could be even more severe, as the source of energy would not be connected to the SMBH located at the GC.
As an example, the rate at which mechanical energy is injected in the CMZ due to supernova explosions is $\approx 10^{40}$~erg~s$^{-1}$ \citep[see][and references therein]{jou17}, which is one order of magnitude smaller than $P_\mathrm{b}$.

We conclude that a source of ionisation of molecular hydrogen in the CMZ other than CRs is very likely to exist.
The most obvious candidate is a radiation field made of UV and/or X-ray photons.
As it is known that X-ray photons emitted by Sgr~A$^*$ during outbursts do not suffice to explain the observed ionisation rates \citep{dog13}, the sources of ionising photons will have to be distributed across the entire CMZ.
Further studies in this direction are therefore needed.

\begin{acknowledgements}
     The authors acknowledge helpful advice on the code from Sebastian-Achim Müller, Francesco Conte, Ludwig M. Böss, Enrico Peretti and Alexandre Inventar. We are also thankful for the inspiring discussions on low-energy CRs in the GC region with Andrea Goldwurm, Marianne Lemoine-Goumard, Denis Allard, Rui-zhi Yang, Richard Tuffs, Jim Hinton, Bing Liu, Adam Ginsburg and Thushara Pillai. This study was supported by the LabEx UnivEarthS, ANR-10-LABX-0023 and ANR-18-IDEX-0001. V.H.M.P. acknowledges support from the Initiative Physique des Infinis (IPI), a research training program of the Idex SUPER at Sorbonne Université.  S.G. acknowledges support from Agence Nationale de la Recherche (grant ANR-21-CE31-0028).
\end{acknowledgements}

\bibliographystyle{aa}
\bibliography{references.bib}

\end{document}